\documentclass[aps]{revtex4}
\usepackage{amsfonts}
\usepackage{amsmath}
\usepackage{amssymb,epsf}
\usepackage{graphics,epsfig,subfigure}

\begin{document}

\title{New perspective for black hole thermodynamics in\\
Gauss-Bonnet--Born-Infeld massive gravity}
\author{Seyed Hossein Hendi$^{1,2}$\footnote{%
email address: hendi@shirazu.ac.ir}, Gu-Qiang
Li$^{3}$\footnote{email address: zsgqli@hotmail.com}, Jie-Xiong
Mo$^{3}$\footnote{email address: mojiexiong@gmail.com}, Shahram
Panahiyan$^{1,4}$ \footnote{email address: sh.panahiyan@gmail.com}
and Behzad Eslam Panah$^{1}$\footnote{email address:
behzad.eslampanah@gmail.com}} \affiliation{$^1$ Physics Department
and Biruni Observatory, College of Sciences, Shiraz
University, Shiraz 71454, Iran\\
$^2$ Research Institute for Astronomy and Astrophysics of Maragha
(RIAAM), P. O. Box 55134-441, Maragha, Iran\\
$^3$ Institute of Theoretical Physics, Lingnan Normal University,
Zhanjiang, 524048, Guangdong, China\\
$^4$ Physics Department, Shahid Beheshti University, Tehran 19839,
Iran}

\begin{abstract}
Following earlier study regarding Einstein-Gauss-Bonnet-massive
black holes in the presence of Born-Infeld nonlinear
electromagnetic field \cite{HendiEPGBBIMass}, we study
thermodynamical structure and critical behavior of these black
holes through different methods in this paper. Geometrical
thermodynamics is employed to give a picture regarding phase
transition of these black holes. Next, a new method is used to
derive critical pressure and horizon radius of these black holes.
In addition, Maxwell equal area law is employed to study the Van
der Waals like behavior of these black holes. Moreover, the
critical exponents are calculated and by using Ehrenfest
equations, the type of phase transitions is determined.
\end{abstract}

\maketitle

\section{Introduction}

Black hole solution is one of the interesting consequences of general
relativity. Although the existence of black holes is vivid, it is an open
question to realize interior nature of them in quantitative detail; the main
reason comes from the fact that a perfect theory of quantum gravity does not
yet exist. Studying the semiclassical phase structure of black holes
provides at least preliminary steps for understanding the quantum gravity.

The phase transition plays an important role for exploring the critical
behavior of the system near critical point. After the discovery of a phase
transition by Hawking and Page \cite{HawkingPage}, black hole phase
transitions have been of great interest. It is known that the asymptotically
flat vacuum black hole solutions are thermally unstable \cite{Dias}. While
asymptotically AdS black holes are the famous examples of the Hawking-Page
phase transition \cite{HawkingPage} between two stable phases. In order to
characterize the critical behavior of a system during the phase transition,
one may calculate its critical exponents which are not completely
independent. Meanwhile two systems belong to the same universality class if
their critical behavior is expressed by the same critical exponents.

The semiclassical phase transition which occurs in the
asymptotically AdS spacetimes can be translated to
confinement/deconfinement phase transition in the context of
AdS/CFT \cite{Witten}. Regarding the applications of AdS/CFT
correspondence in recent years, the similarities between the phase
transition of black holes and holographic superconductivity, have
achieved a great deal of attention
\cite{Supperconductor1,Supperconductor2}.

The local (thermal) stability of a system is concerned with how the system
responds to small fluctuations of its thermodynamic coordinates. There are
various methods that one can employ to investigate the phase structure of a
black hole system near its critical point. One of the well-known standard
analysis of the locally stability is based on canonical ensemble by studying
the specific heats. Phase structure may be also explained by critical
quantities that are extracted in the extended phase space. In addition, one
may apply the geometrical thermodynamics method for studying the phase
transition.

One of the methods for constructing phase structure of a
thermodynamical system is through the use of geometry. Meaning, by
considering a thermodynamical potential and its corresponding
extensive parameters, it is possible to introduce a metric which
describes thermodynamical properties of the system. The
information regarding thermodynamical properties of the system is
extracted from Ricci scalar of the metric. The divergencies of
Ricci scalar of the thermodynamical metric mark two important
points of the thermodynamical system; bound point and phase
transition. There are several
methods regarding thermodynamical geometry which are; Weinhold \cite%
{WeinholdI,WeinholdII}, Ruppeiner \cite{RuppeinerI,RuppeinerII},
Quevedo \cite{QuevedoI,QuevedoII} and HPEM
\cite{HPEMI,HPEMII,HPEMIII}. The geometrical thermodynamics has
been employed in the context of different types of black holes
\cite{HanC,BravettiMMA,Ma,Mo2014,GarciaMC,ZhangCY,MoLW2016}. In
addition, this method was also used to study phase transition of
superconductors \cite{BasakCNS}. A comparative study regarding
different geometrical thermodynamical metrics is done in Ref.
\cite{comp}. A successful method of the geometrical thermodynamics
include all bound and phase transition points in its Ricci scalar
through the divergencies. In other words, divergencies of the
Ricci scalar and mentioned points must coincide with each other. A
mismatch and extra divergency indicate the existence of anomaly
which contradict with principles of thermodynamics. Such anomaly
was reported for Weinhold, Ruppeiner and Quevedo metrics for
different types of black holes
\cite{HPEMI,HPEMII,HPEMIII,MassiveBTZ}. To overcome such problem,
HPEM metric was introduced \cite{HPEMI}. In this paper, we will
regard HPEM method for studying geometrical thermodynamics of
black holes under consideration.

Einstein gravity introduces gravitons as massless particles,
whereas there are several arguments that state gravitons should be
massive particles. In order to have massive graviton, theory of
general relativity should be modified to include mass terms. The
first attempt for constructing a massive theory was referred to
the works of Fierz and Pauli \cite{Fierz} which was done in the
context of linear theory. This theory has specific problem which
is known as van Dam, Veltman and Zakharov discontinuity. Meaning
that propagator of the massive gravity in limit $m=0$ is not
consistent with one derived for massless case. The resolution to
this problem was Vainshtein mechanism which requires the system to
be considered in nonlinear regime. In other words, according to
the Vainshtein mechanism \cite{VainshteinA}, at some distance
below the so-called Vainshtein radius, the linear regime breaks
down and the theory enters into a nonlinear framework. Based on
this mechanism, the usual general relativity can be recovered from
high curvature space-times which are introduced with a wide class
of non-Einsteinian theories. In the context of a static and
spherically symmetric base space, it is also shown that the
Vainshtein mechanism can work, correctly, both inside and outside
the compact objects \cite{VainshteinB,VainshteinC} (See Refs.
\cite{Vainshtein1,Vainshtein2,Vainshtein3,Vainshtein4,Vainshtein5}
for more details regarding Vainshtein mechanism). On the other
hand, generalization of the Fierz and Pauli massive theory to
nonlinear one leads to existence of a Boulware-Deser ghost
\cite{Boulware}. While solutions to
these problems had existed for some time in three dimensional spacetime \cite%
{DeserJ,Bergshoeff}, they were not solved in four and higher dimensions. In
order to solve such problems de Rham, Gabadadze and Tolley (dRGT) proposed
another class of massive gravity \cite{dRGTI,dRGTII}. Contrary to previous
theories, dRGT theory is valid in higher dimensions and it was shown that
such theory enjoys absence of the Boulware-Deser ghost \cite%
{HassanR,HassanRS}. This theory build up massive terms by employing a
reference metric. A modification in reference metric could lead to another
dRGT like massive theory \cite{Vegh}. Black hole solutions of dRGT massive
gravity and their thermodynamical properties have been investigated for $d$%
-dimensions ($d\geq 3$) in Refs. \cite%
{MassiveBTZ,CaiHPZ,HendiMassive1,HendiMassive2,XuCH,Ghosh,Do}.

On the other hand, one of the well-known theories of higher derivative
gravity is Lovelock theory which is a natural generalization of Einstein
gravity in higher dimensions. Taking into account the first additional term
of Einstein gravity in the context of Lovelock theory (Gauss-Bonnet (GB)
gravity), it is believed that GB gravity can solve some of the shortcomings
of Einstein gravity \cite{Stelle,Maluf,Farhoudi}. In addition, GB gravity
consists curvature-squared terms which, interestingly, is free of ghosts and
the corresponding field equations contain no more than second derivatives of
the metric (see Refs. \cite{BoulwareD,Zumino,Myers,Callan,Cho,Cai} for more
details). Another interesting aspect of GB gravity is that it can be arisen
from the low-energy limit of heterotic string theory \cite%
{Zwiebach,GrossW,Metsaev,MetsaevT}. Considering GB gravity
context, black hole solutions and their interesting behavior have
been investigated in many literatures
\cite{ChoN,CaiG,Barrau,Dotti,Charmousis,HendiE,HendiPM,Maselli,Dadhich}.

On the other hand, one of the main problems of Maxwell's
electromagnetic field theory for a point-like charge is that there
is a singularity at the charge position and therefore, it has
infinite self energy. In order to remove this self energy, in
classical electrodynamics, Born and Infeld introduced a nonlinear
electromagnetic field \cite{BornI}, with main motivation of
solving the infinite self energy problem by imposing a maximum
strength of the electromagnetic field. Motivated by the
interesting results mentioned above, we study thermodynamic
behavior of black holes in GB-massive gravity in the present
Born-Infeld (BI) source.

In order to have a better description regarding physics governing
a system, it is necessary to decrease different shortcomings of
different theories as much as it is possible. This indicates that
we should apply more generalizations to solve different
shortcomings of theories describing the nature of system. Here, we
have considered three generalizations; BI generalization to remove
shortcomings of the Maxwell theory, GB gravity to solve different
problems of Einstein theory such as renormalization problem, and
massive gravity to solve the massless gravitons in both Einstein
gravity and GB theory. Such considerations solve some of the
shortcomings of theories under consideration, but they also modify
the physical properties of the system. In this paper, we intend to
investigate these modifications in the context of critical
behavior of black holes which in turn provides a reasonable
framework for conducting studies in other aspects of physics such
as gauge/gravity duality.

The outline of the paper will be as follow. In next section, we
introduce action and basic equations related to GB-BI-massive
gravity. We also present a brief discussion regarding to the black
hole solutions and conserved and thermodynamics quantities.
Section \ref{GTs} is devoted to study the phase transition through
geometrical thermodynamics. In Sec. \ref{PV}, we investigate
critical behavior of the system via a new method, which comes from
the maximum point of denominator of heat capacity. We also check
the Maxwell equal area law in Sec. \ref{MEAL}. After that we
calculate the critical exponents of the system in the extended
phase space in Sec. \ref{CritExp}. In Sec. \ref{Ehrenfest}, we
examine the Ehrenfest equations at the critical point and confirm
the validity of second order phase transition. In the last section
we present our conclusions.

%%%%%%%%%%%%%%%%%%%%%%%%%%%%%%%%%%%%%%%%%%%%%%%%%%%%%%%%%%%%%

\section{Basic Equations}

In the current paper, we set out to discuss the geometric and
thermodynamic properties of charged black holes in $d$-dimensional
GB-massive gravity with $d-4$ compact dimensions. Regarding
compactified extra dimensions, it has been shown that, depending
on the horizon topology, one can obtain black string/membrane
solutions in addition to black hole solutions. Furthermore, it has
been pointed out that, in the context of GB gravity, one may
obtain non-trivial modified solutions with an extra asymptotic
charge \cite{modified}. In this paper, we focus on the black hole
solutions with usual conserved charges.

The $d$-dimensional action of GB-massive gravity with the negative
cosmological constant and in the presence of BI electrodynamics is
\begin{eqnarray}
\mathcal{I} &=&-\frac{1}{16\pi }\int d^{d}x\sqrt{-g}\left[ \mathcal{R}%
-2\Lambda +\alpha \left( R_{\mu \nu \gamma \delta }R^{\mu \nu \gamma \delta
}-4R_{\mu \nu }R^{\mu \nu }+R^{2}\right) \right.   \notag \\
&&\left. +4\beta ^{2}\left( 1-\sqrt{1+\frac{\mathcal{F}}{2\beta ^{2}}}%
\right) +m^{2}\sum_{i}^{4}c_{i}\mathcal{U}_{i}(g,f)\right] ,  \label{Action}
\end{eqnarray}%
where $\mathcal{R}$, $\Lambda $, $m$, $\alpha $ and $\beta $ are,
respectively, the scalar curvature, the cosmological constant, the massive
parameter, the GB factor and BI parameter. Also $R_{\mu \nu }$ and $R_{\mu
\nu \gamma \delta }$ are Ricci and Riemann tensors, $\mathcal{F}=F_{\mu \nu
}F^{\mu \nu }$ denotes the Maxwell invariant and $f$ is a fixed symmetric
tensor. In Eq. (\ref{Action}), $c_{i}$'s are constants and $\mathcal{U}_{i}$%
's are symmetric polynomials of the eigenvalues of $d\times d$ matrix $%
\mathcal{K}_{\nu }^{\mu }=\sqrt{g^{\mu \alpha }f_{\alpha \nu }}$, which can
be written in the following forms
\begin{eqnarray}
\mathcal{U}_{1} &=&\left[ \mathcal{K}\right] , \\
\mathcal{U}_{2} &=&\left[ \mathcal{K}\right] ^{2}-\left[ \mathcal{K}^{2}%
\right] , \\
\mathcal{U}_{3} &=&\left[ \mathcal{K}\right] ^{3}-3\left[ \mathcal{K}\right] %
\left[ \mathcal{K}^{2}\right] +2\left[ \mathcal{K}^{3}\right] , \\
\mathcal{U}_{4} &=&\left[ \mathcal{K}\right] ^{4}-6\left[ \mathcal{K}^{2}%
\right] \left[ \mathcal{K}\right] ^{2}+8\left[ \mathcal{K}^{3}\right] \left[
\mathcal{K}\right] +3\left[ \mathcal{K}^{2}\right] ^{2}-6\left[ \mathcal{K}%
^{4}\right] .
\end{eqnarray}

Using the action (\ref{Action}) and variation of this action with
respect to the metric tensor ($g_{\mu \nu }$) and the Faraday
tensor ($F_{\mu \nu }$), respectively, lead to
\begin{equation}
G_{\mu \nu }+\Lambda g_{\mu \nu }+H_{\mu \nu }-\frac{1}{2}g_{\mu \nu }L(%
\mathcal{F})-\frac{2F_{\mu \lambda }F_{\nu }^{\lambda }}{\sqrt{1+\frac{%
\mathcal{F}}{2\beta ^{2}}}}+m^{2}\chi _{\mu \nu }=0,  \label{Field equation}
\end{equation}%
\begin{equation}
\partial _{\mu }\left( \frac{\sqrt{-g}F^{\mu \nu }}{\sqrt{1+\frac{\mathcal{F}%
}{2\beta ^{2}}}}\right) =0,  \label{Maxwell equation}
\end{equation}%
in the above equations $G_{\mu \nu }$ is the Einstein tensor, $H_{\mu \nu }$
and $\chi _{\mu \nu }$ are
\begin{eqnarray}
H_{\mu \nu } &=&-\frac{\alpha }{2}\left[ 8R^{\rho \sigma }R_{\mu \rho \nu
\sigma }-4R_{\mu }^{\rho \sigma \lambda }R_{\nu \rho \sigma \lambda
}-4RR_{\mu \nu }+8R_{\mu \lambda }R_{\nu }^{\lambda }+\right.   \notag \\
&&\left. g_{\mu \nu }\left( R_{\mu \nu \gamma \delta }R^{\mu \nu \gamma
\delta }-4R_{\mu \nu }R^{\mu \nu }+R^{2}\right) \right] ,  \label{GBterm}
\end{eqnarray}
and
\begin{eqnarray}
\chi _{\mu \nu } &=&-\frac{c_{1}}{2}\left( \mathcal{U}_{1}g_{\mu \nu }-%
\mathcal{K}_{\mu \nu }\right) -\frac{c_{2}}{2}\left( \mathcal{U}_{2}g_{\mu
\nu }-2\mathcal{U}_{1}\mathcal{K}_{\mu \nu }+2\mathcal{K}_{\mu \nu
}^{2}\right) -\frac{c_{3}}{2}(\mathcal{U}_{3}g_{\mu \nu }-3\mathcal{U}_{2}%
\mathcal{K}_{\mu \nu }  \notag \\
&&+6\mathcal{U}_{1}\mathcal{K}_{\mu \nu }^{2}-6\mathcal{K}_{\mu \nu }^{3})-%
\frac{c_{4}}{2}(\mathcal{U}_{4}g_{\mu \nu }-4\mathcal{U}_{3}\mathcal{K}_{\mu
\nu }+12\mathcal{U}_{2}\mathcal{K}_{\mu \nu }^{2}-24\mathcal{U}_{1}\mathcal{K%
}_{\mu \nu }^{3}+24\mathcal{K}_{\mu \nu }^{4}).  \label{MassiveTerm}
\end{eqnarray}

\subsection{Black hole solutions}

Considering the metric of $d$-dimensional spacetime as
\begin{equation}
ds^{2}=-f(r)dt^{2}+f^{-1}(r)dr^{2}+r^{2}h_{ij}dx_{i}dx_{j},\
i,j=1,2,3,...,n~,  \label{Metric}
\end{equation}%
where $h_{ij}dx_{i}dx_{j}$ is the line element with constant curvature $%
\left( d-2\right) (d-3)\kappa $ and volume $V_{d-2}$ and the ansatz metric
in the following form \cite{Vegh}%
\begin{equation}
f_{\mu \nu }=diag(0,0,c^{2}h_{ij}),  \label{f11}
\end{equation}%
in which $c$ is a positive constant. The metric function was
obtained in Ref. \cite{HendiEPGBBIMass} as
\begin{eqnarray}
f\left( r\right) &=&\kappa +\frac{r^{2}}{2\alpha d_{3}d_{4}}\left\{ 1-\sqrt{%
1+\frac{8\alpha d_{3}d_{4}}{d_{1}d_{2}}\left[ \Lambda +\frac{d_{1}d_{2}m_{0}%
}{2r^{d_{1}}}+\mathcal{A}+\mathcal{B}\right] }\right\} ,  \label{f(r)} \\
\mathcal{A} &=&-2\beta ^{2}\left( 1-\sqrt{1+\eta }\right) -\frac{d_{2}^{2} q^{2}}{%
r^{2d_{2}}}\mathcal{H},  \notag \\
\mathcal{B} &=&-m^{2}d_{1} d_{2} \left[ \frac{d_{3}d_{4}c^{4}c_{4}}{%
2r^{4}}+\frac{d_{3}c^{3}c_{3}}{2r^{3}}+\frac{c^{2}c_{2}}{2r^{2}}+\frac{cc_{1}%
}{2d_{2}r}\right] ,  \notag
\end{eqnarray}%
where $m_{0}$ and $q$ are integration constants which are related
to the total mass and the electric charge of black hole,
respectively. The notation $d_i$ is introduced by us to denote the
term $d-i$ (Recall that $d$ is the spacetime dimensionality) so as
to simplify the expressions of physical quantities in this paper.
For example, $d_1$ denotes the term $d-1$ while $d_2$ denotes the
term $d-2$. It is notable that, in the above solution, we used the
gauge potential ansatz $A_{\mu }=h(r)\delta _{\mu }^{0}$ in the
Maxwell equation (\ref{Maxwell equation}). Also, $\mathcal{H}$,
$\eta$ and consistent $h(r)$ are in the following forms
\begin{eqnarray}
\mathcal{H} &=&{}_{2}F_{1}\left( \left[ \frac{1}{2},\frac{d_{3}}{2d_{2}}%
\right] ,\left[ \frac{3d_{7/3}}{2d_{2}}\right] -\eta \right) , \\
\eta &=&\frac{d_{2}d_{3}q^{2}}{2\beta ^{2}r^{2d_{2}}}, \\
h(r)&=&-\sqrt{\frac{d_{2}}{2d_{3}}}\frac{q}{r^{d_{3}}} \mathcal{H}.
\end{eqnarray}

It was shown that the asymptotical behavior of the solutions are
(a)dS solutions with an effective cosmological constant
($\Lambda_{eff}$) \cite{HendiEPGBBIMass}. This effective
cosmological constant reduces to ordinary $\Lambda$ for vanishing
$\alpha$. It was also shown that neither massive nor BI parts
affect the asymptotical behavior of the solutions
\cite{HendiEPGBBIMass}.

\subsection{Thermodynamics}

The Hawking temperature of the black hole is given by
\cite{HendiEPGBBIMass}
\begin{eqnarray}
T &=&\frac{1}{4\pi \mathcal{N}}\left\{ \frac{m^{2}}{r_{+}}\left[
d_{3}d_{4}\left( c^{3}c_{3}r_{+}+d_{5}c^{4}c_{4}\right) +r_{+}^{2}\left(
cc_{1}r_{+}+d_{3}c^{2}c_{2}\right) \right] +\frac{2r_{+}^{3}}{d_{2}}\left(
2\beta ^{2}-\Lambda \right) \right.  \notag \\
&&\left. -\frac{4\beta ^{2}r_{+}^{3}}{d_{2}\Upsilon _{+}}+\frac{\kappa d_{3}%
}{r_{+}}\left( r_{+}^{2}+\alpha \kappa d_{4}d_{5}\right) \right\} ,
\label{TotalTT}
\end{eqnarray}%
where $\mathcal{N}=2\alpha \kappa d_{3}d_{4}+r_{+}^{2}$ and also, $\Upsilon
_{+}=\Upsilon \left\vert _{r=r_{+}}\right. $ (which $\Upsilon =\sqrt{%
1-\left( \frac{h^{\prime }(r)}{\beta }\right) ^{2}}$). It is
notable that $r_+$ in the above expression denotes the largest
real root of equation $f(r)=0$.

The total charge, the electric potential ($U$) and the entropy of the black
hole are \cite{HendiEPGBBIMass}
\begin{eqnarray}
Q &=&\frac{V_{d_{2}}\ \sqrt{d_{2}d_{3}}}{4\pi }q,  \label{TotalQ} \\
U &=&A_{\mu }\chi ^{\mu }\left\vert _{r\rightarrow \infty }\right. -A_{\mu
}\chi ^{\mu }\left\vert _{r\rightarrow r_{+}}\right. =\sqrt{\frac{d_{2}}{%
2d_{3}}}\frac{q}{r_{+}^{d_{3}}}\ \mathcal{H}_{+},  \label{TotalU} \\
S &=&\frac{V_{d_{2}}}{4}r_{+}^{d_{2}}\left( 1+\frac{2d_{2}d_{3}}{r_{+}^{2}}%
\kappa \alpha \right) ,  \label{TotalS}
\end{eqnarray}%
where $\mathcal{H}_{+}=\mathcal{H}\left\vert _{r=r_{+}}\right. $.
Total mass of the black hole is in the following form
\cite{HendiEPGBBIMass}
\begin{equation}
M=\frac{\ d_{2}\ V_{d_{2}}}{16\pi }m_{0}.  \label{TotalM}
\end{equation}

The first law of thermodynamics for black hole solution in the GB-BI-massive
gravity was checked in Ref. \cite{HendiEPGBBIMass} and it was found that
these thermodynamics quantities satisfy the first law of black hole
thermodynamics as
\begin{equation}
dM=TdS+UdQ.
\end{equation}

\section{geometrical thermodynamics \label{GTs}}

Here, we are interested in studying the critical behavior of the
black holes through the use of geometrical method. This method
builds phase space of black holes by using one of the
thermodynamical quantities as thermodynamical potential and its
corresponding extensive parameters as components of phase space.
By doing so, a metric is obtained in which thermodynamical
properties of the system are stored in its Ricci scalar.
Divergencies of the Ricci scalar point out two important places in
thermodynamical behavior of the system; whether system goes under
second order phase transition or it meets a bound point. A bound
point is where heat capacity/temperature meets a root. In other
words, in bound points a limit for having physical system
(positive temperature) is given. On the other hand, in phase
transition point, heat capacity has a divergency, implying that
there is a discontinuity in heat capacity. In place of this
divergency, a second order phase transition takes place.

\begin{figure}[tbp]
$%
\begin{array}{cccc}
\epsfxsize=4.5cm \epsffile{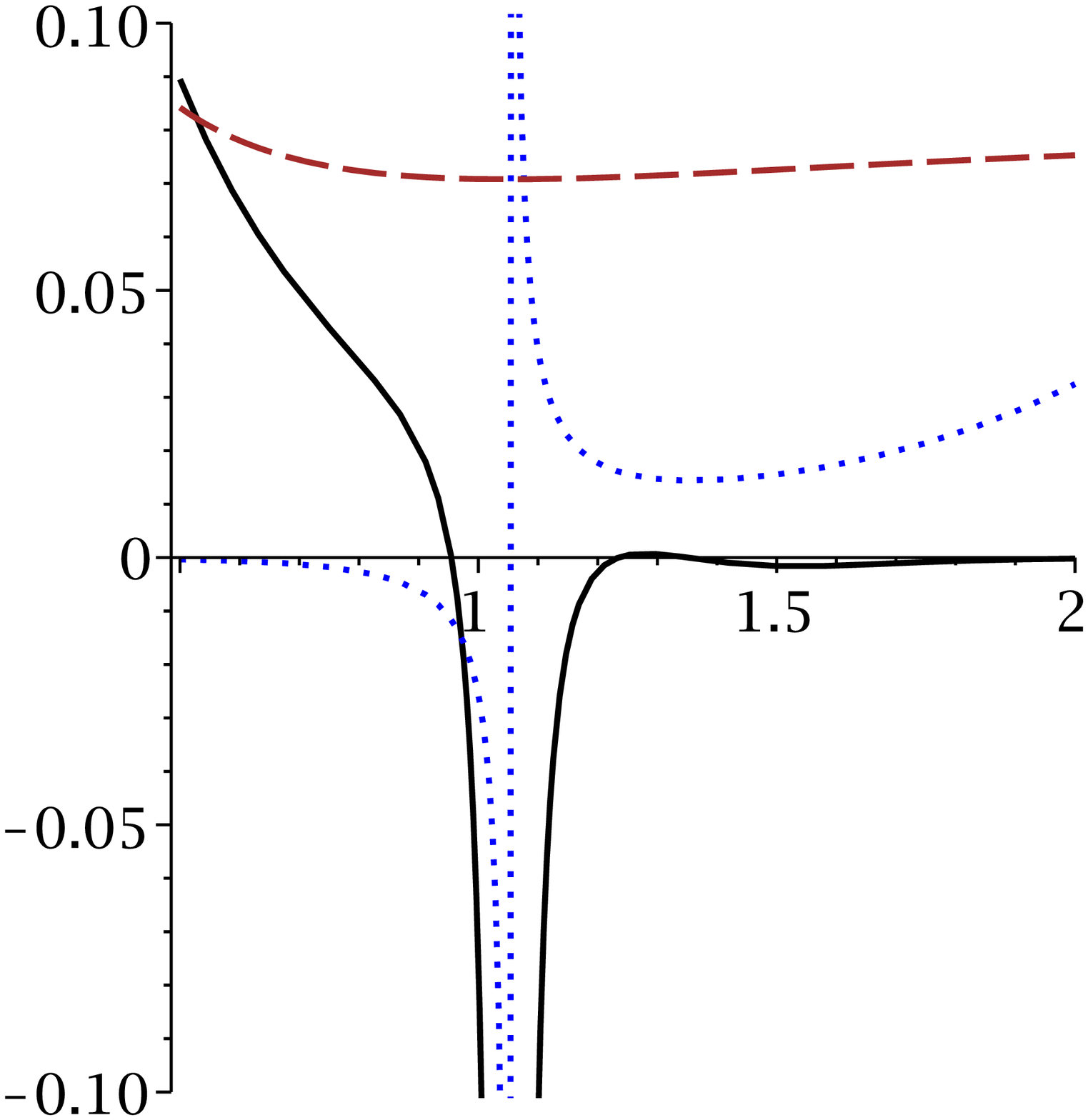} & \epsfxsize=4.5cm %
\epsffile{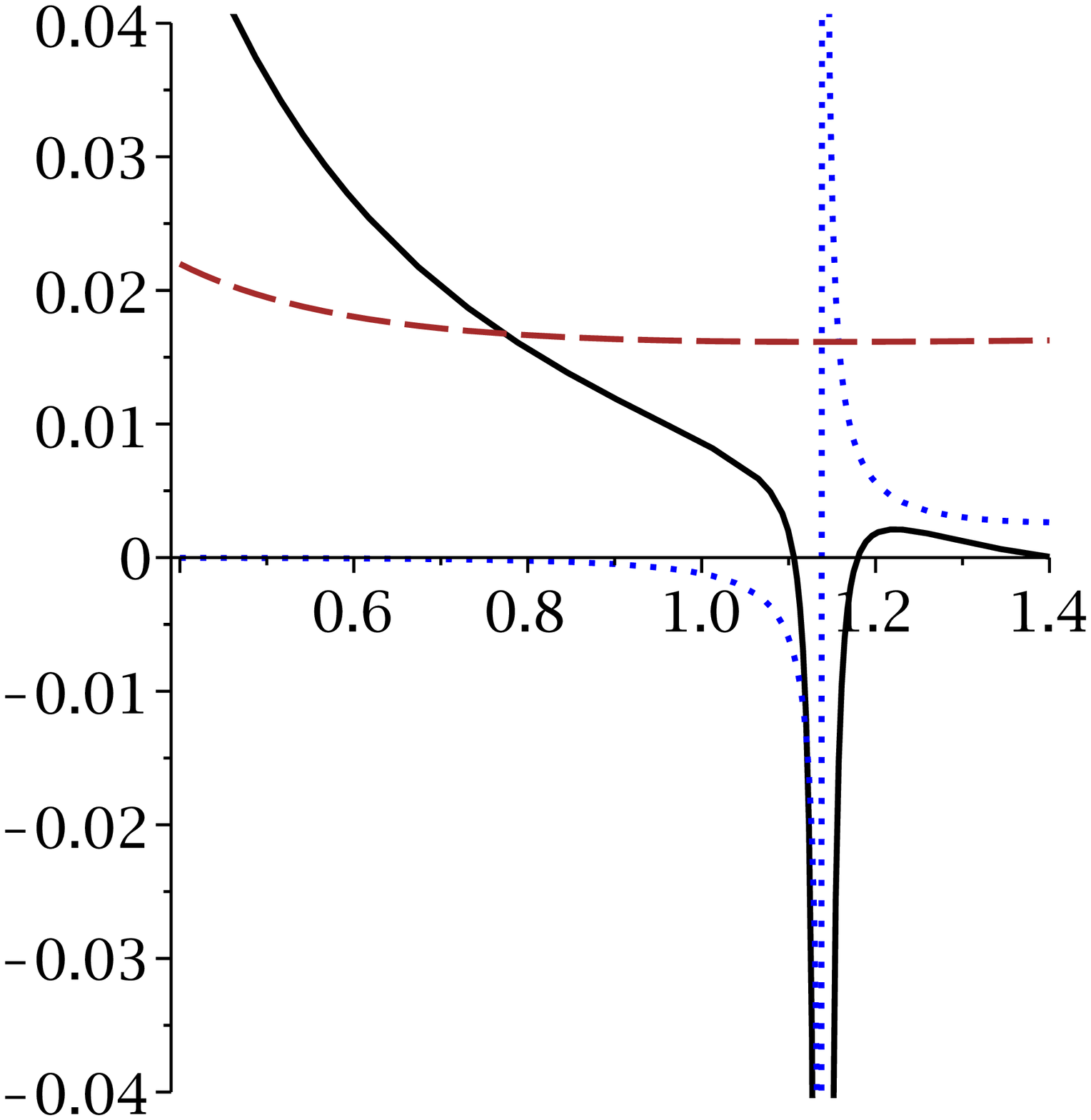} & \epsfxsize=4.5cm \epsffile{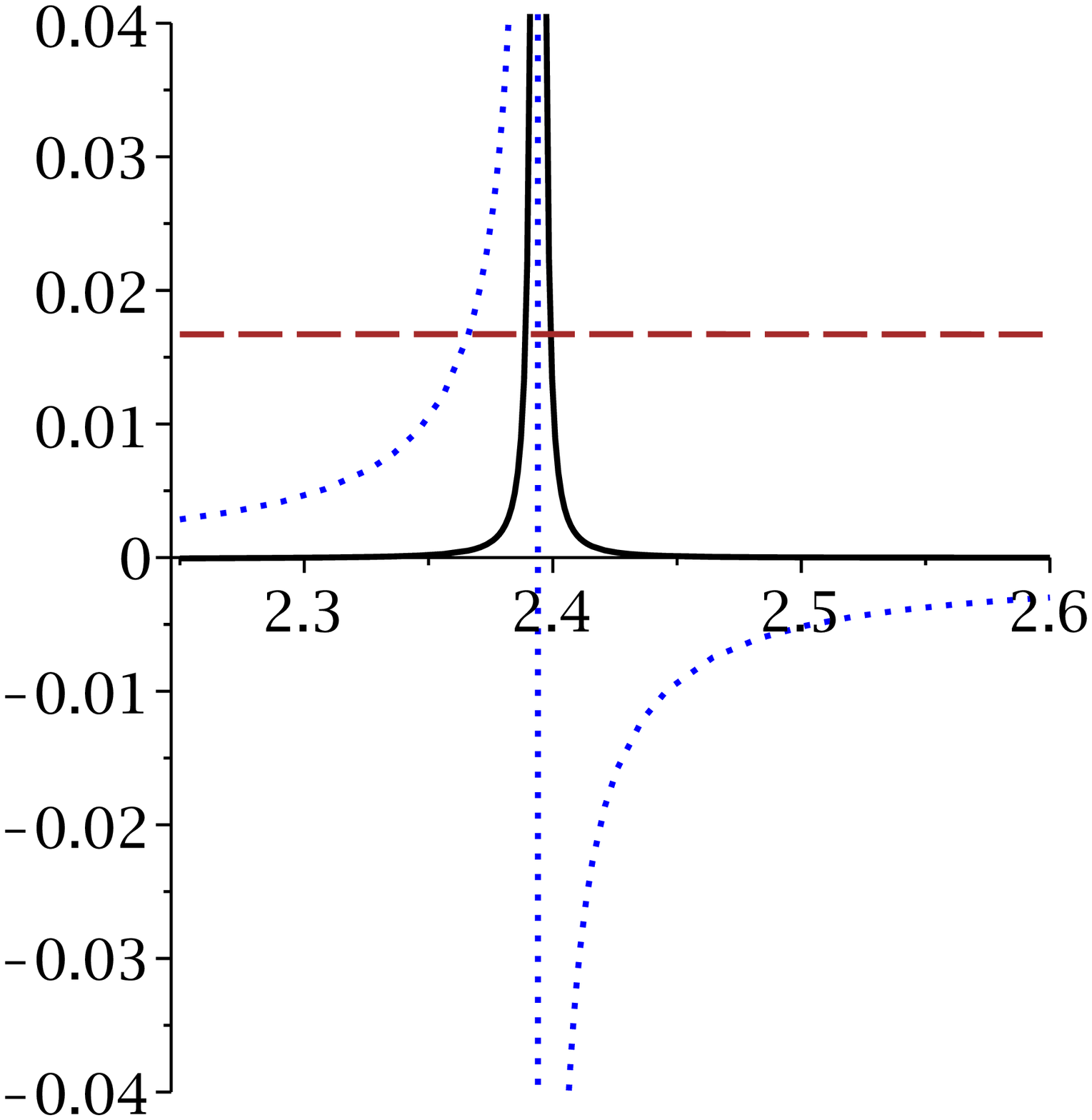} & %
\epsfxsize=4.5cm \epsffile{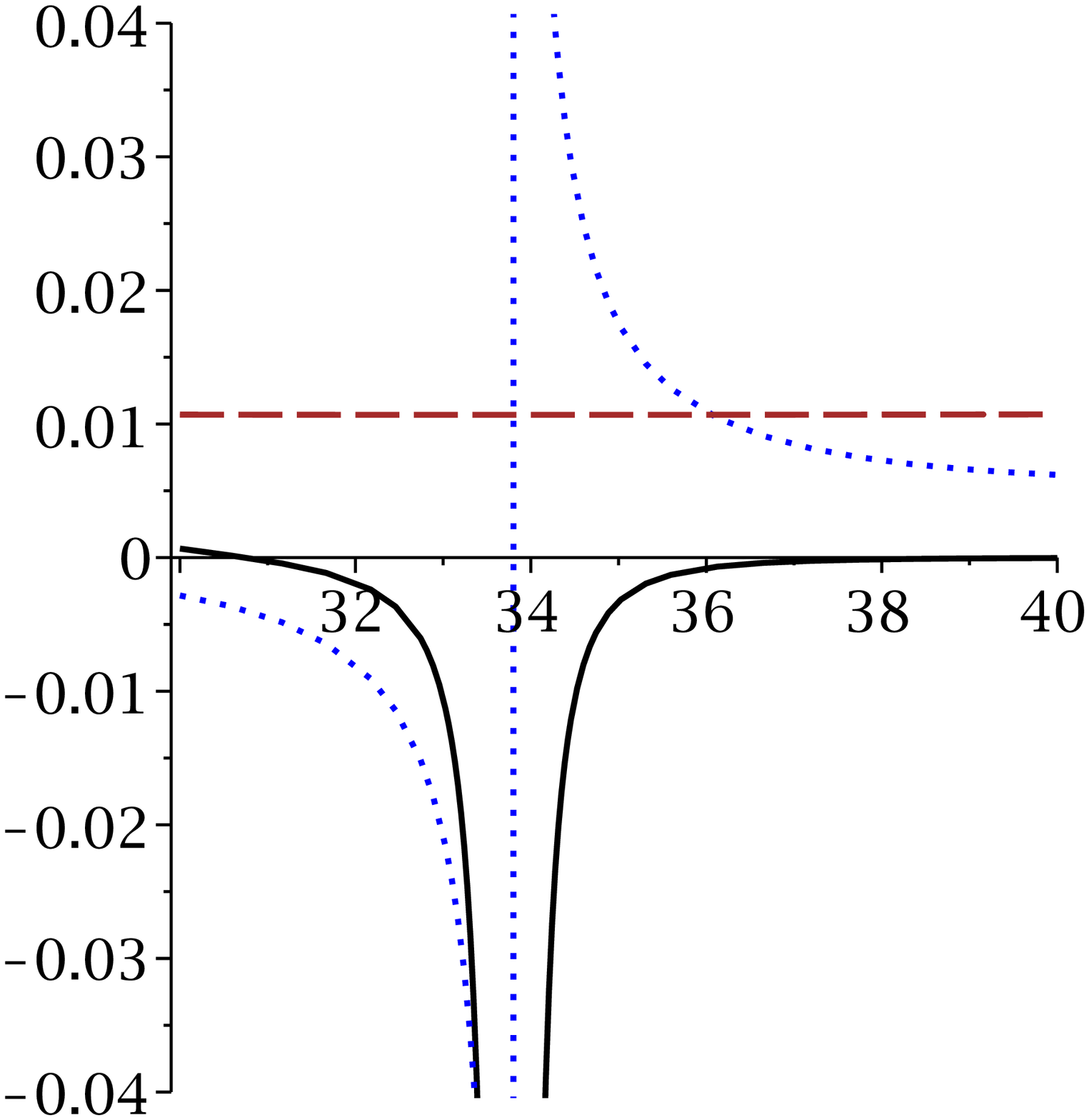}%
\end{array}
$%
\caption{For different scales: $\mathcal{R}$ (continuous line),
$C_{Q}$ (dotted line) and $T$ (dashed line) versus $r_{+}$ for
$q=1$, $\Lambda =-1$,
$c=c_{1}=c_{2}=2$, $c_{3}=c_{4}=0.2$, $k=1$, $\protect\beta=0.5$, $d=6$ and $%
\protect\alpha=0.5$; left panel: $m=1$; three right panels: $m=5$.}
\label{Fig1}
\end{figure}

\begin{figure}[tbp]
$%
\begin{array}{cccc}
\epsfxsize=4.5cm \epsffile{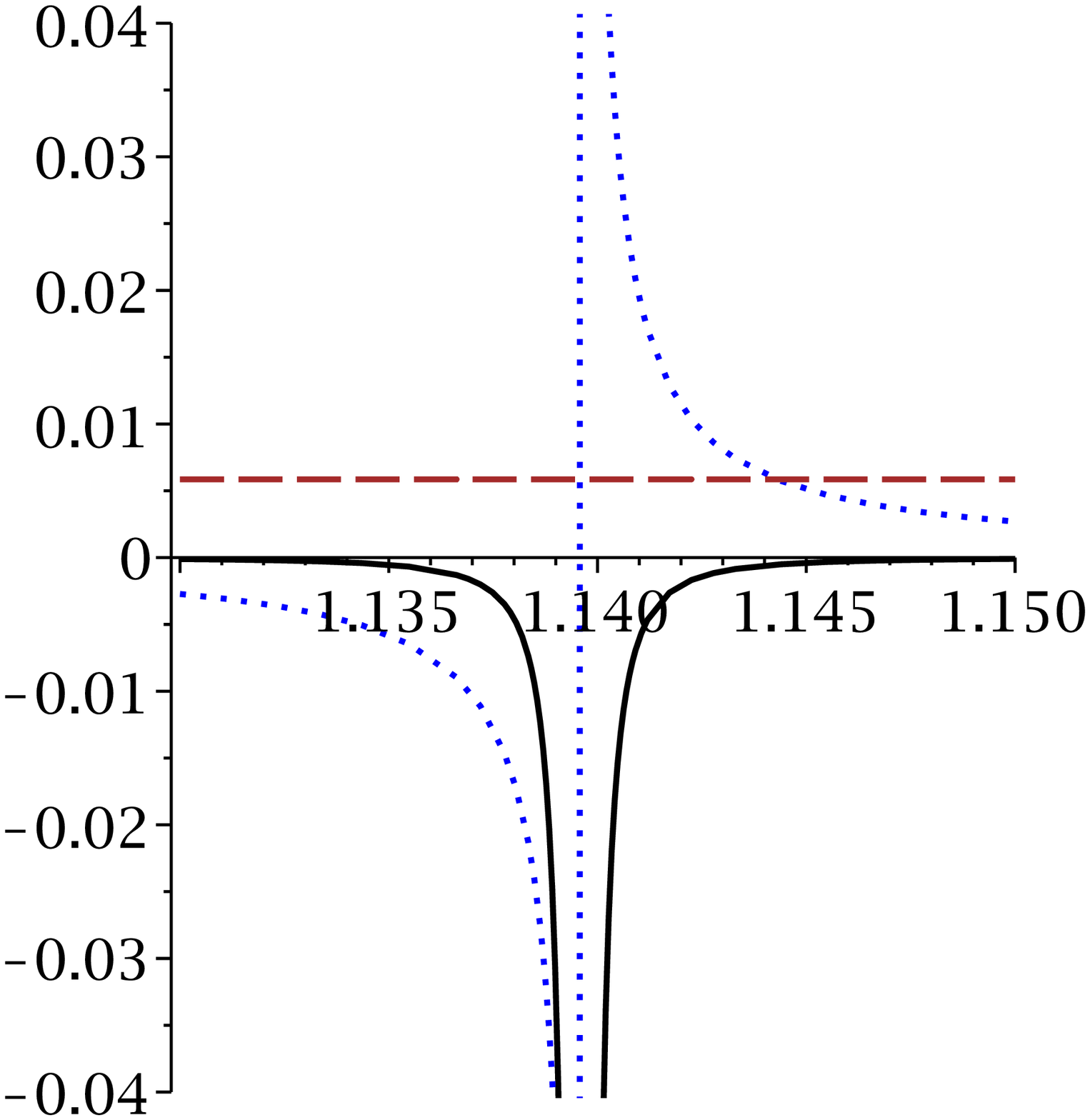} & \epsfxsize=4.5cm %
\epsffile{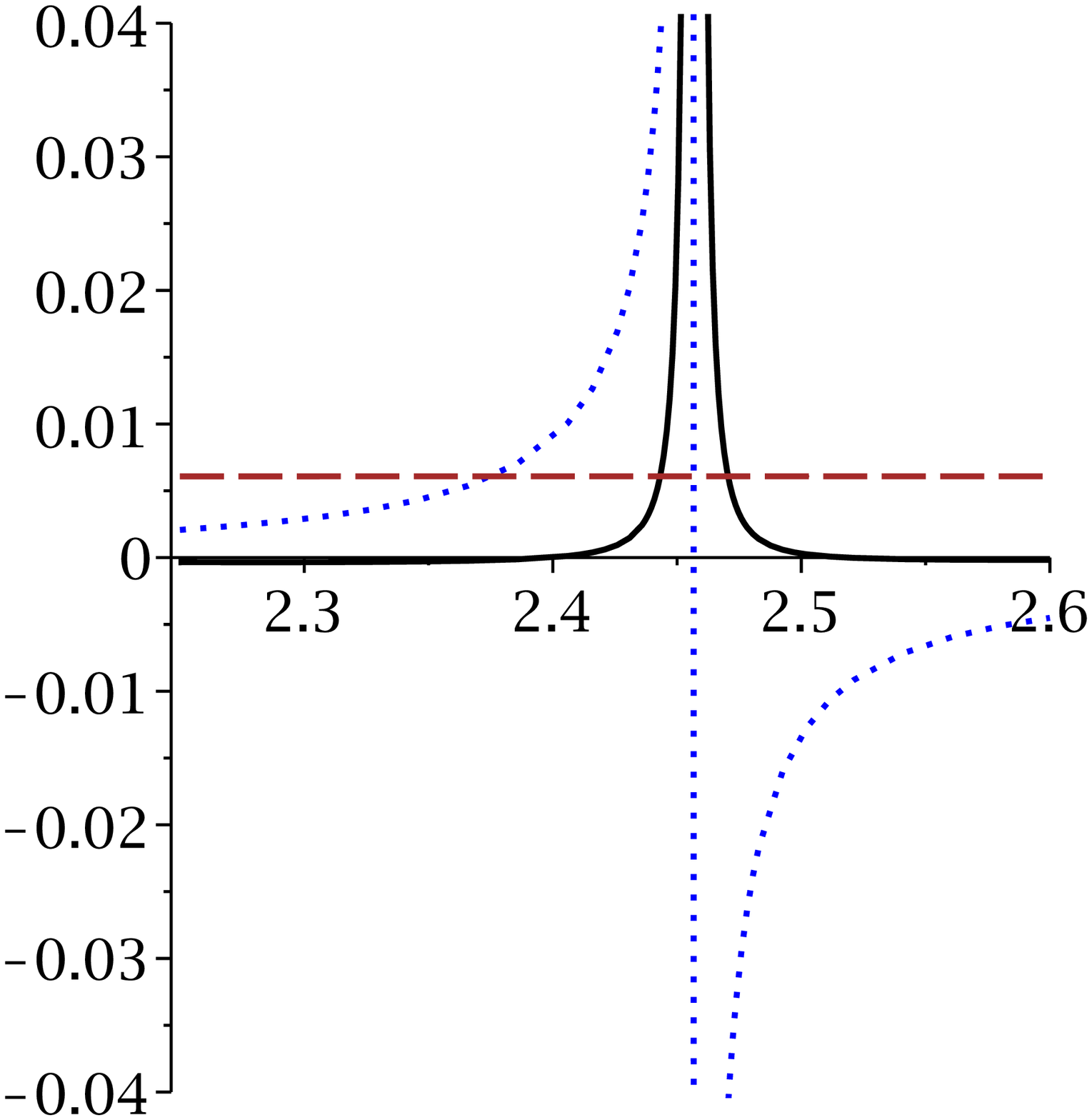} & \epsfxsize=4.5cm %
\epsffile{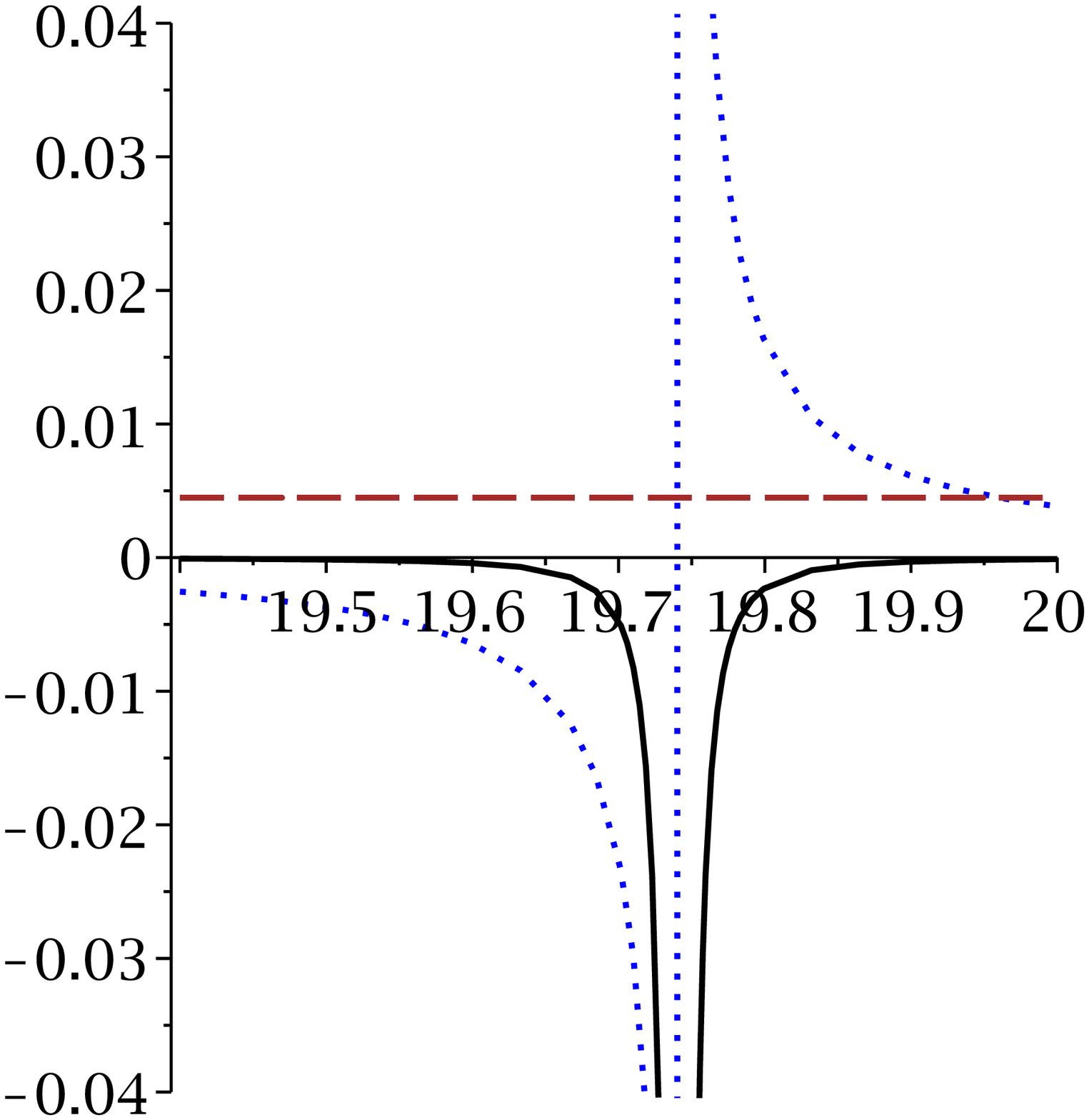} &  \\
\epsfxsize=4.5cm \epsffile{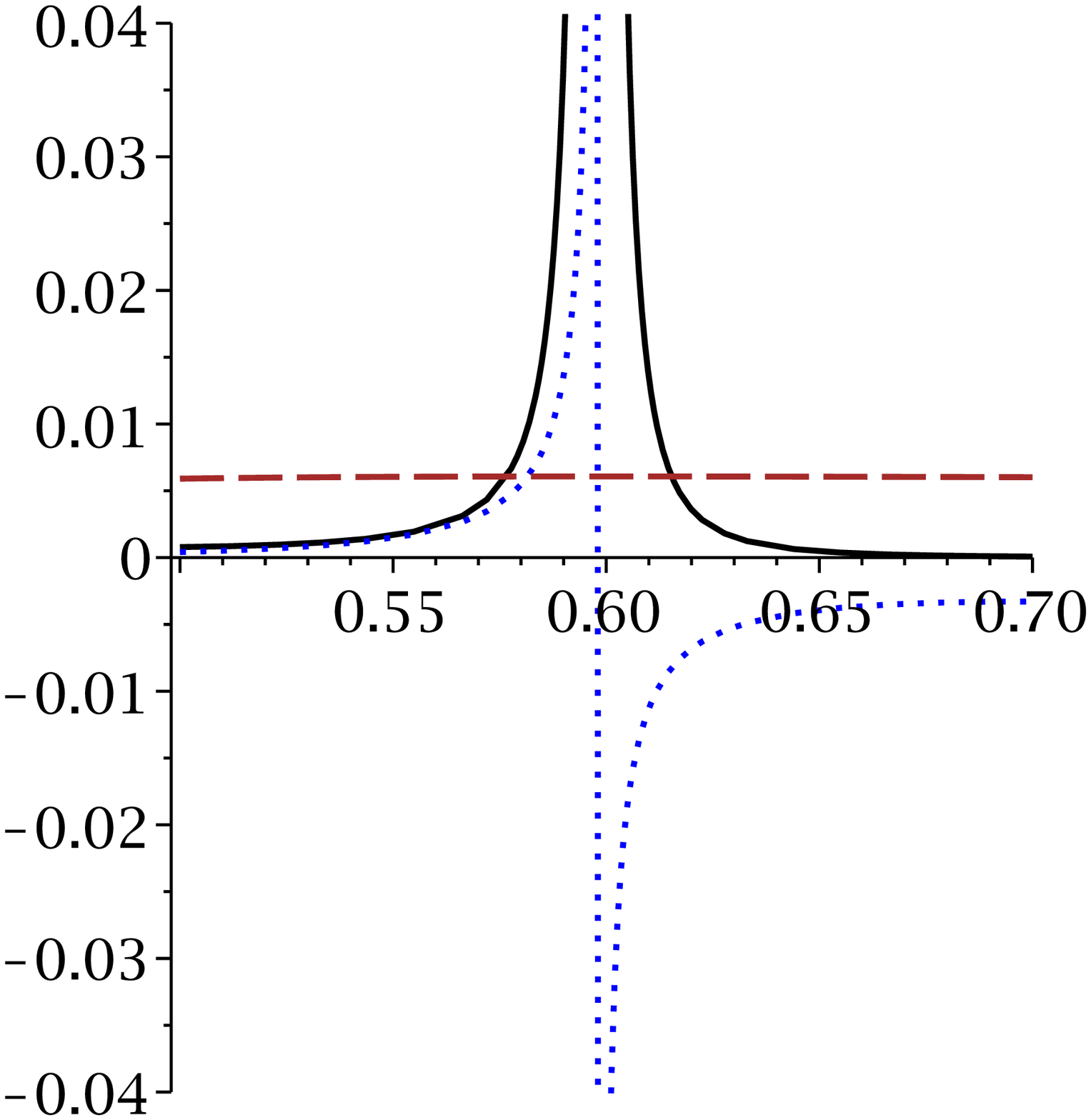} & \epsfxsize=4.5cm %
\epsffile{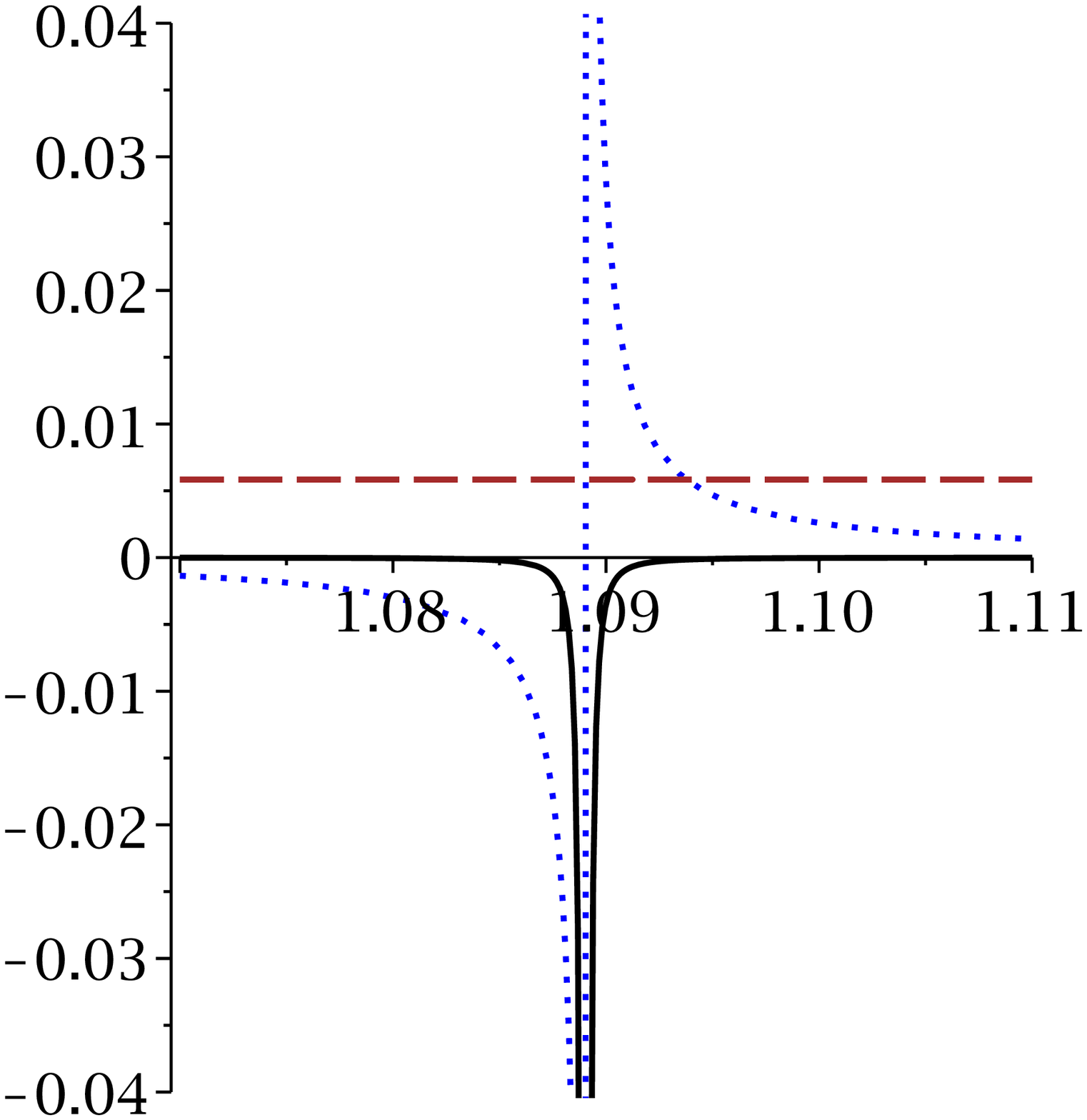} & \epsfxsize=4.5cm %
\epsffile{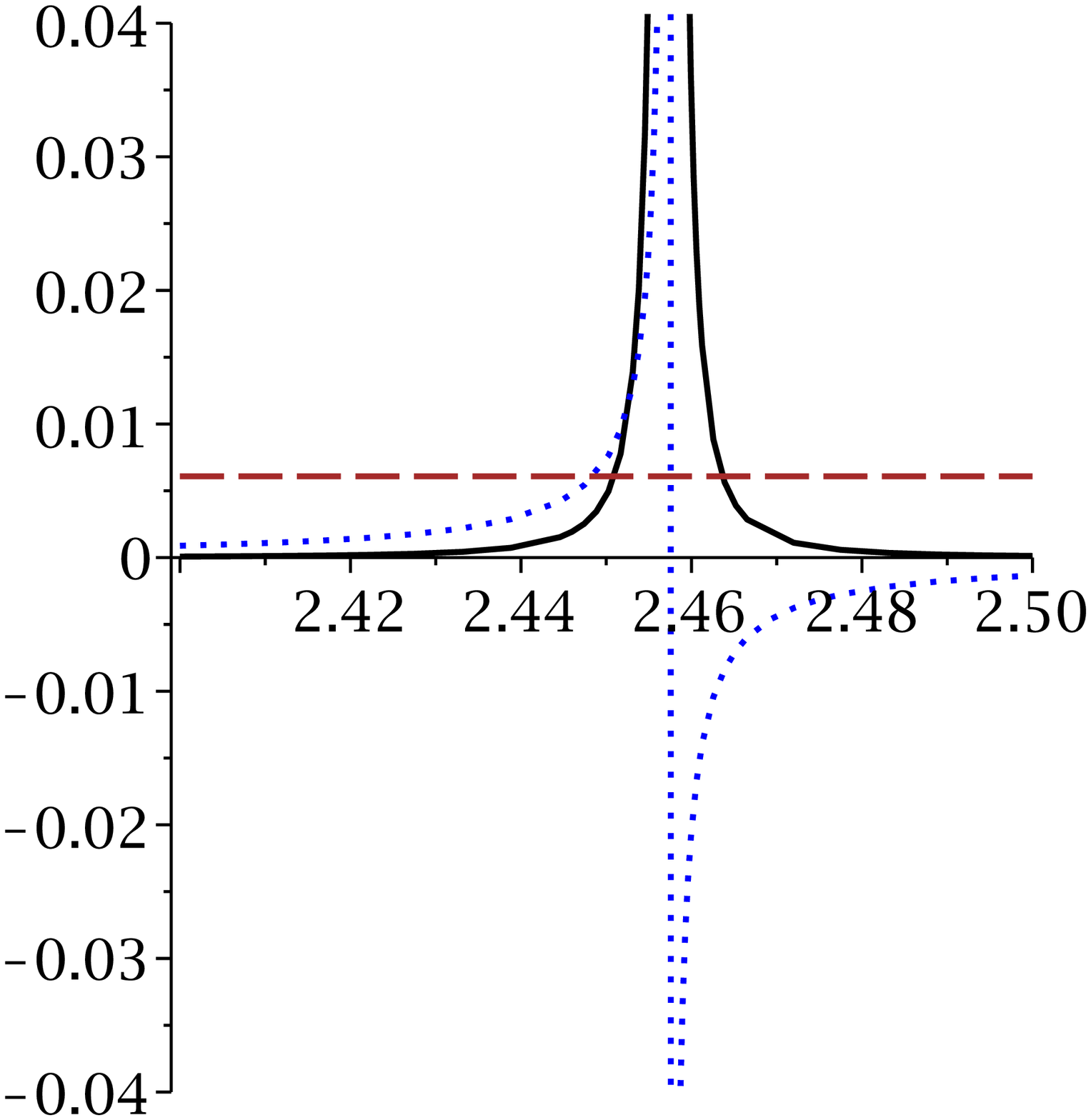} & \epsfxsize=4.5cm %
\epsffile{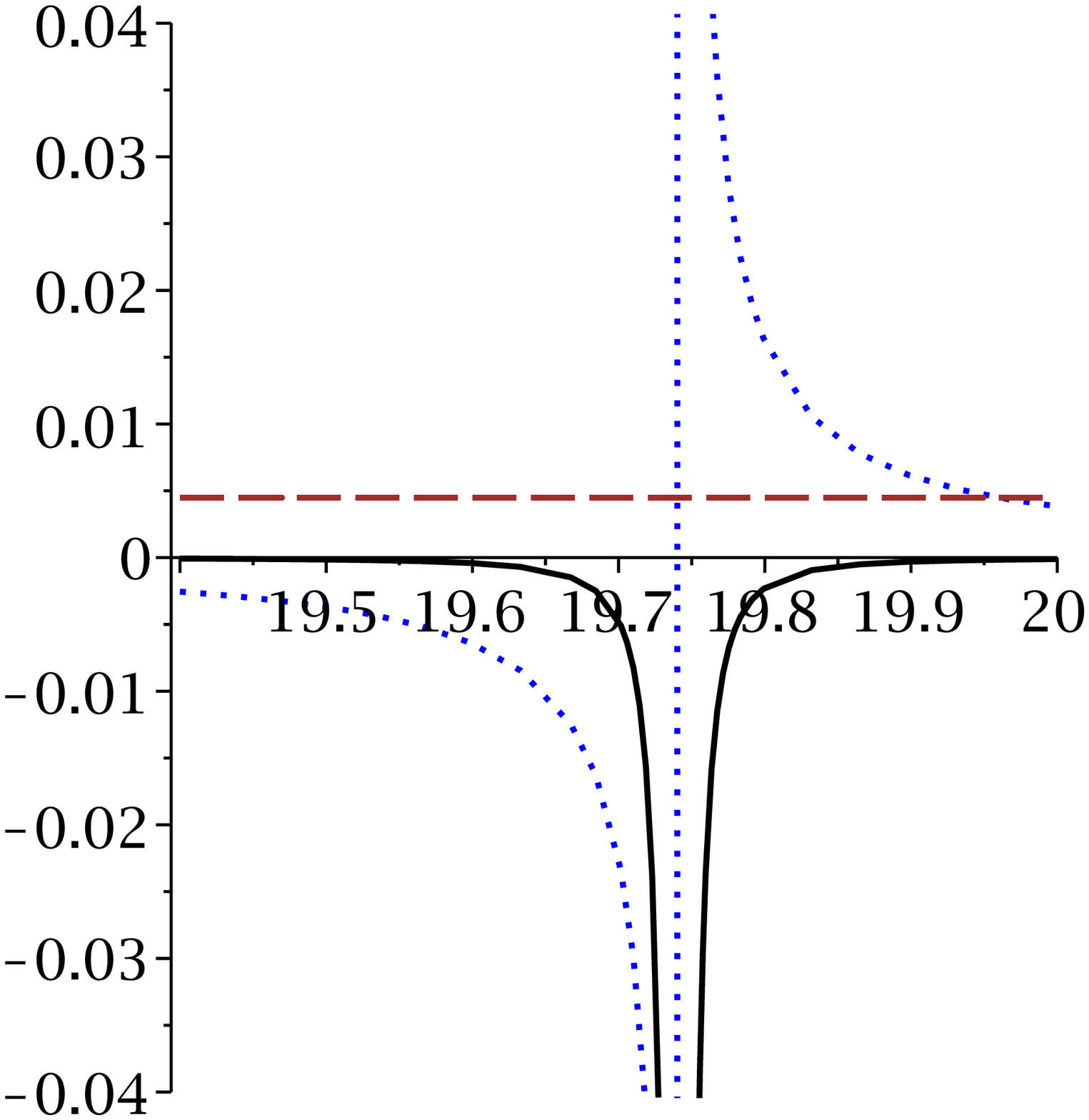}%
\end{array}
$%
\caption{For different scales: $\mathcal{R}$ (continuous line),
$C_{Q}$ (dotted line) and $T$ (dashed line) versus $r_{+}$ for
$q=1$, $\Lambda =-1$,
$c=c_{1}=c_{2}=2$, $c_{3}=c_{4}=0.2$, $k=1$, $m=3$, $d=6$ and $\protect\alpha%
=0.5$; up panels: $\protect\beta=0.1$; down panels: $\protect\beta=100$.}
\label{Fig2}
\end{figure}

\begin{figure}[tbp]
$%
\begin{array}{cccc}
\epsfxsize=4.5cm \epsffile{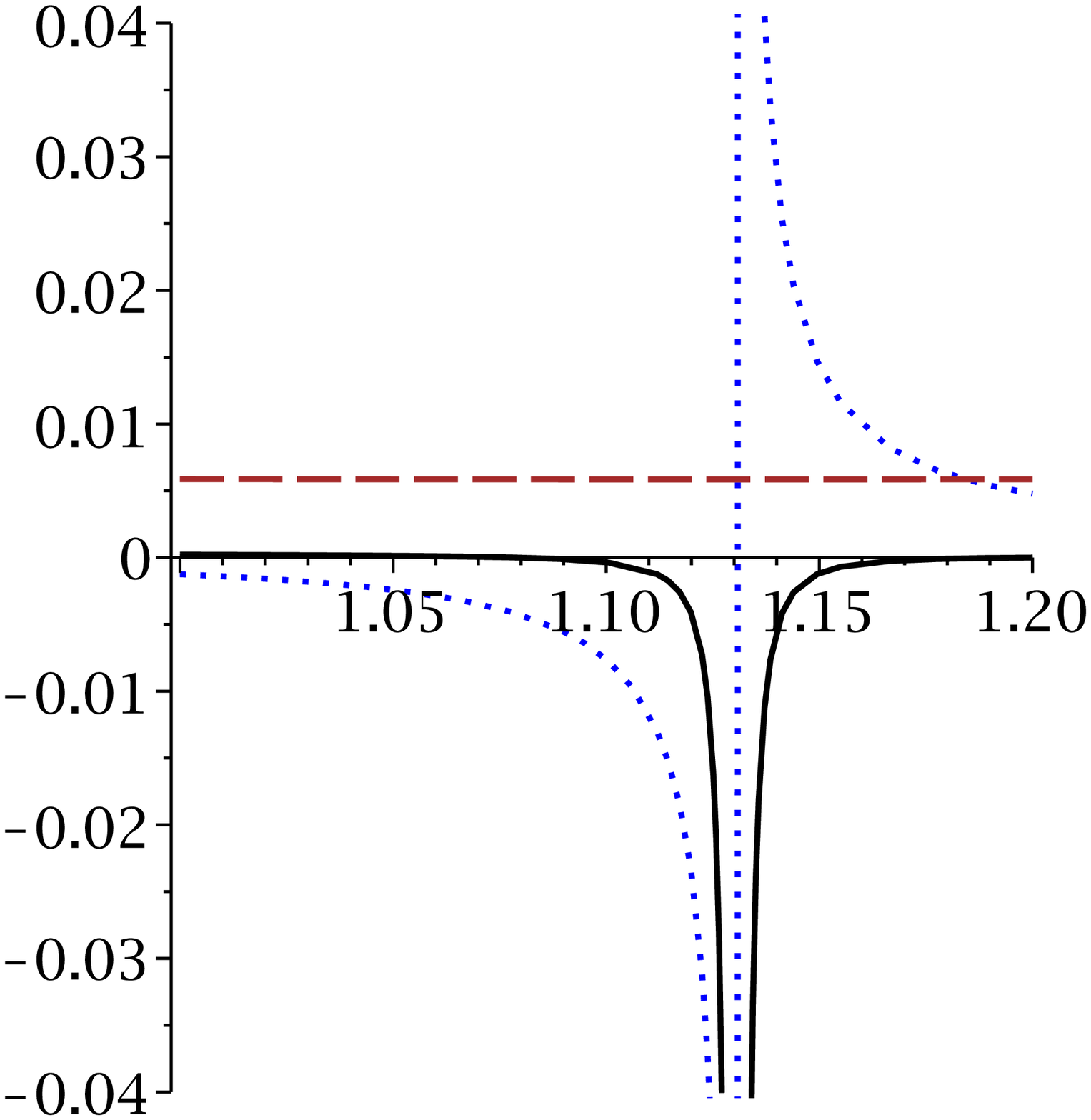} & \epsfxsize=4.5cm %
\epsffile{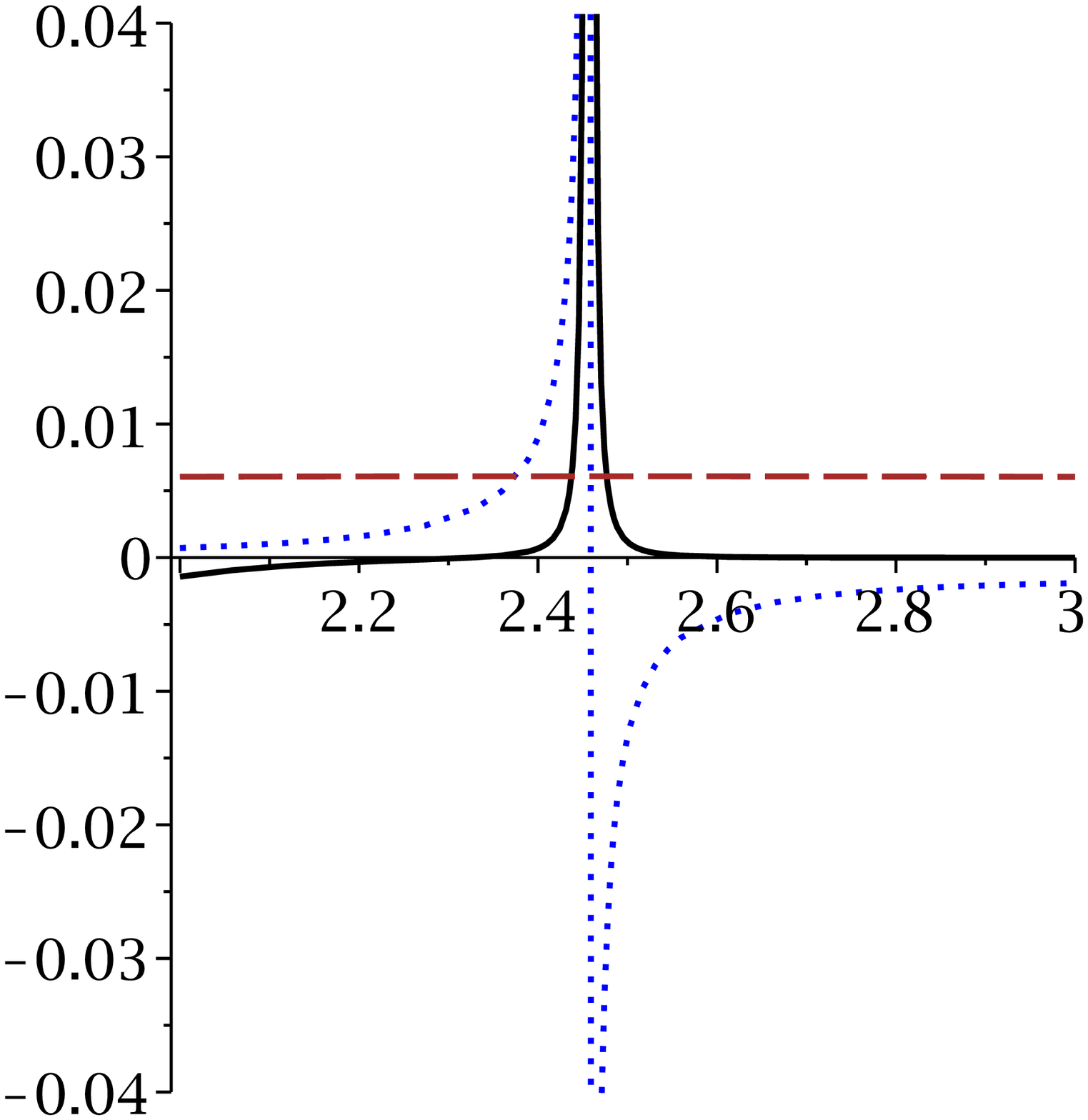} & \epsfxsize=4.5cm %
\epsffile{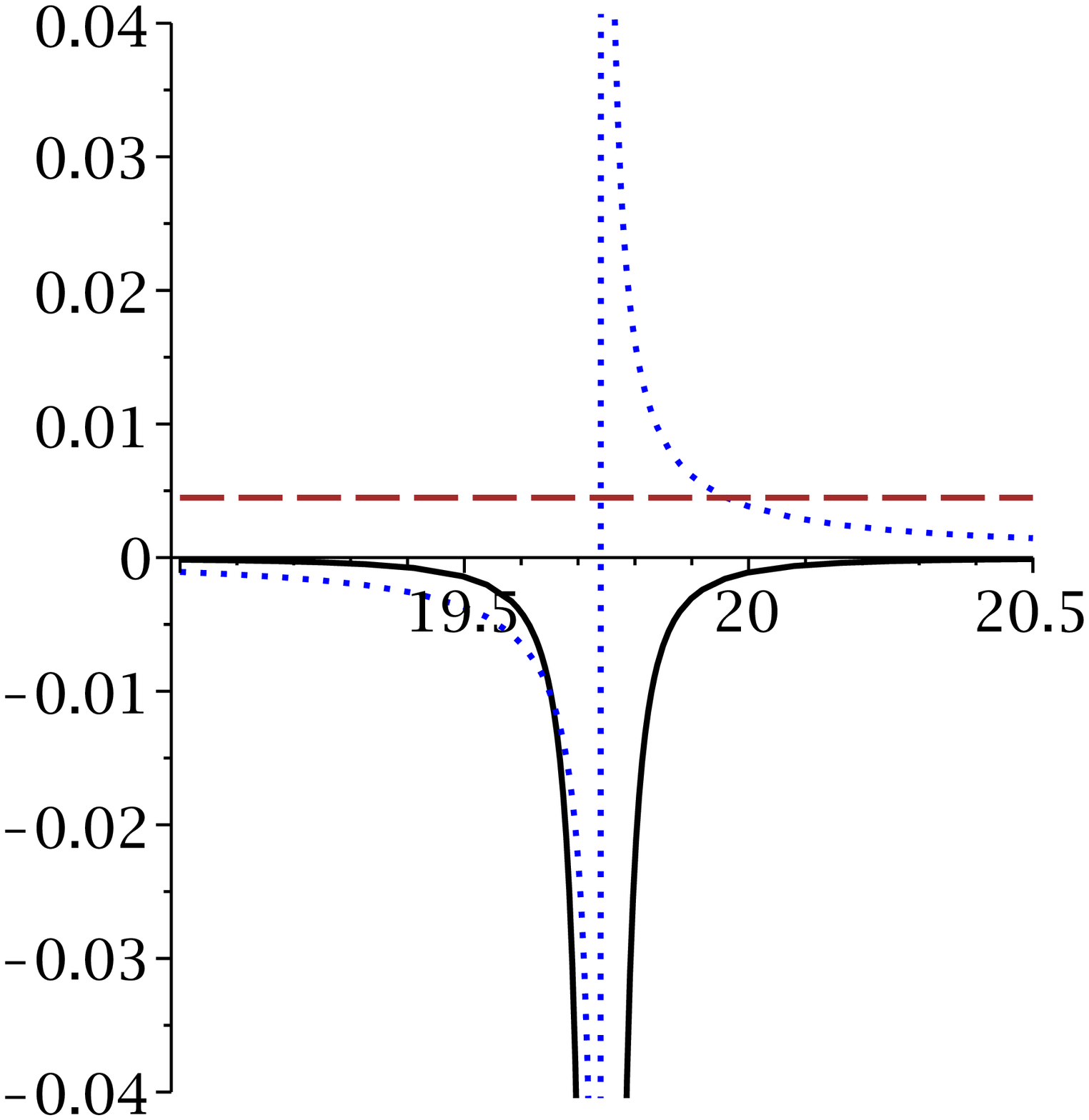} & \epsfxsize=4.5cm %
\epsffile{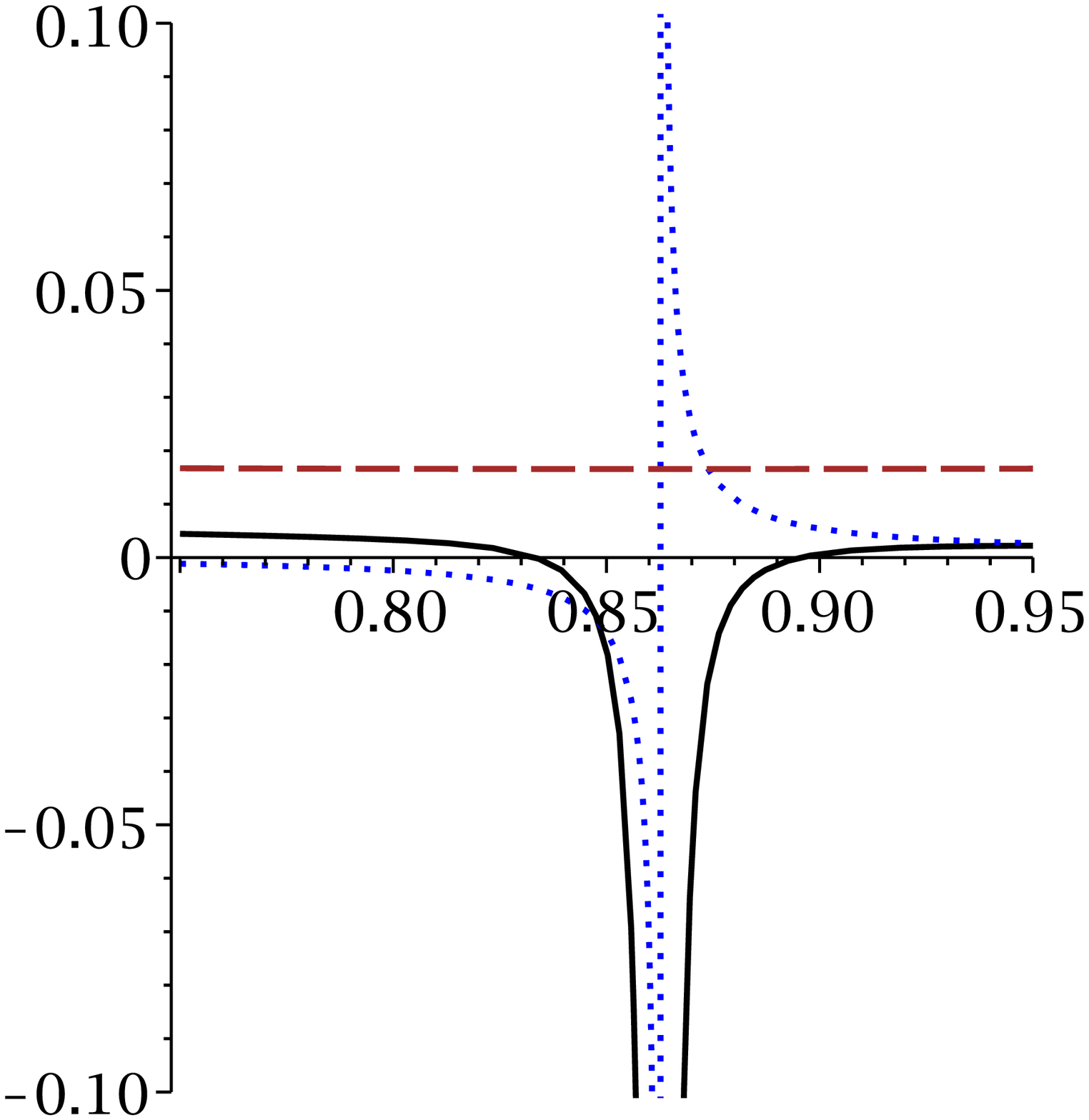}%
\end{array}
$%
\caption{For different scales: $\mathcal{R}$ (continuous line),
$C_{Q}$ (dotted line) and $T$ (dashed line) versus $r_{+}$ for
$q=1$, $\Lambda =-1$,
$c=c_{1}=c_{2}=2$, $c_{3}=c_{4}=0.2$, $k=1$, $\protect\beta=0.5$, $d=6$ and $%
m=3$; three left panels: $\protect\alpha=0.5$; right panel: $\protect\alpha%
=2 $.}
\label{Fig3}
\end{figure}

There are several methods for constructing phase space of black
holes through thermodynamical quantities; Weinhold
\cite{WeinholdI,WeinholdII}, Ruppeiner
\cite{RuppeinerI,RuppeinerII}, Quevedo \cite{QuevedoI,QuevedoII}
and HPEM \cite{HPEMI,HPEMII,HPEMIII}. A successful method should
cover all the mentioned points without any extra divergency for
its Ricci scalar. Existence of extra divergency or mismatch
between divergency of Ricci scalar and phase transition (or bound
points) indicate that there is a case of anomaly. Recently, it was
shown that employing Weinhold, Ruppeiner and Quevedo may lead to
existence of anomaly \cite{HPEMI,HPEMII,HPEMIII}. To overcome the
problems of other methods, HPEM metric was proposed. The structure
of HPEM metric is
\begin{equation}
ds^{2}= S\frac{M_{S}}{M_{QQ}^{3}}\left( -M_{SS}dS^{2}+M_{QQ}dQ^{2}\right),
\end{equation}%
where $M_{X}=\partial M/\partial X$ and $M_{XX}=\partial ^{2}M/\partial
X^{2} $. Now, by using total mass of black holes (\ref{TotalM}) with entropy
(\ref{TotalS}) and electric charge (\ref{TotalQ}), one can construct phase
space and calculate its Ricci scalar. Due to economical reasons, we will not
present obtained Ricci scalar but rather present its results in following
diagrams (Figs. \ref{Fig1}-\ref{Fig3}).

%%%%%%%%%%%%%%%%%%%%%%%%%%%%%%%%%%%%%%%%%%%%%%%%%%%%%%%%%%%%%%%

%%%%%%%%%%%%%%%%%%%%%%%%%%%%%%%%%%%%%%%%%%%%%%%%%%%%%%%%%%%%%%%
%%%%%%%%%%%%%%%%%%%%%%%%%%%%%%%%%%%%%%%%%%%%%%%%%%%%%%%%%%%%%%%

%%%%%%%%%%%%%%%%%%%%%%%%%%%%%%%%%%%%%%%%%%%%%%%%%%%%%%%%%%%%%%%
%%%%%%%%%%%%%%%%%%%%%%%%%%%%%%%%%%%%%%%%%%%%%%%%%%%%%%%%%%%%%%%

%%%%%%%%%%%%%%%%%%%%%%%%%%%%%%%%%%%%%%%%%%%%%%%%%%%%%%%%%%%%%%%

Evidently, the number of phase transition points and their places
are functions of massive (Fig. \ref{Fig1}), BI (Fig. \ref{Fig2})
and GB (Fig. \ref{Fig3}) parameters. For considered values of
different parameters, these black holes enjoy the absence of bound
point. In other words, for all values of the horizon radius,
physical black holes exist. On other other hand, these black holes
have second order phase transition in their thermodynamical
structure. The number of these phase transitions may vary from one
(see left panel of Fig. \ref{Fig1} and right panel of Fig.
\ref{Fig3}) to several (see Fig. \ref{Fig2}) phase transitions.

The system has positive temperature but depending on the choices
of different parameters, temperature may acquire one to several
extrema. These extrema are where the heat capacity meets
divergency. In other words, extrema of the temperature are places
in which the heat capacity is divergent. Therefore, these extrema
are places in which black holes go under the second order phase
transition. The number of divergencies in the heat capacity,
hence, phase transitions, is an increasing function of the massive
(see Fig. \ref{Fig1}) and BI (see Fig. \ref{Fig2}) parameters
while it is a decreasing function of the GB parameter (see Fig.
\ref{Fig3}).

Regarding BI theory, for large values of the nonlinearity
parameter, system behaves like Reissner-Nordstr\"{o}m black hole.
Meaning for large values of this parameter, the effect of
nonlinearity decreases and system behaves like it is in the
presence of Maxwell theory of electromagnetic field. On the other
hand, for small values of nonlinearity parameter, system has
Schwarzschild like behavior. Taking these limits into account, one
can extract following conclusions: The highest number of phase
transitions, hence, the highest complexity in phase structure of
these black holes is acquired for linear electromagnetic field.
The generalization to nonlinear electromagnetic field reduces the
number of phase transition and it may omit some of the phase
transitions. By increasing power of the nonlinearity (decreasing
the nonlinearity parameter), system would obtain the least number
of phase transition which acquirable for charged black holes in
this theory of the nonlinear electromagnetic field.

The GB gravity is a higher order gravity. In other words, value of
the Ricci scalar, which is a parameter to measure curvature of the
system, is higher in this theory of gravity comparing to Einstein
gravity. Therefore, the gravity in this theory is stronger
comparing to Einstein theory of gravity. The GB gravity provides
an extra degree of freedom in term of GB parameter. Increasing GB
parameter leads to increasing the value of Ricci scalar, hence,
power of the gravity. We see for these black holes, that by
increasing GB parameter, the number of phase transitions
decreases. Meaning that for system with higher power of gravity
(larger curvature), the number of phase transition and complexity
in phase structure of these black holes decrease. Therefore,
gravity here has an opposing effect on the number of phase
transition.

Massive parameter is directly related to the mass of graviton.
Plotted diagrams for variation of the massive parameter show that
as the mass of graviton increases the black holes under
consideration go under more phase transitions. In other words, by
increasing the mass of graviton the complexity in thermal behavior
and phase structure of these black holes increase.

It is evident that using HPEM metric provides suitable
divergencies in its Ricci scalar for phase transitions that are
observed in the heat capacity. In other words, divergences of the
Ricci scalar of HPEM metric coincide with the phase transition
points of heat capacity. Therefore, these two approaches yield
consistent results. On the other hand, depending on the type of
phase transition (smaller to larger or larger to smaller black
holes), the sign of divergency of the Ricci scalar differs. If the
transition is from larger to smaller, the sign of the Ricci scalar
around the corresponding transition is positive, while the
opposite (negative sign) is observed for the transition of smaller
to larger black holes. These two differences in sign enable one to
determine the type of phase transition of a system.

\section{$P-V$ criticality through new approach \label{PV}}

In this section, we will regard critical behavior of these black
holes through the use of a new method which was introduced in Ref.
\cite{int}. This method employs denominator of the heat capacity
of black holes to extract a relation for pressure. This relation
is independent from equation of state. The maximum of obtained
relation is where phase transition takes place. In other words,
the pressure and horizon radius of maximum of this relation is
where system goes under the second order phase transition and Van
der Waals like behavior is observed. In addition, the picture that
this method draws for pressure smaller/larger than critical
pressure is consistent with thermodynamical behavior for the
system with same pressure in usual thermodynamical systems. In
other words, for pressures smaller than critical pressure, two
horizon radii exist, which marks two different phases in phase
diagrams while for a pressure larger than critical pressure, no
phase transition is observed. This is consistent with behavior of
the $T-V$ diagrams in which for pressures larger than critical
pressure, no phase transition region exists. The consistency of
this method with other methods was
investigated in several papers \cite%
{HendiMassive1,HendiMassive2,HendiMassive3,HendiRainbow}.

Now, we will employ this method to obtain the critical pressure
and horizon radius of these black holes. First, we use the
proportionality between the cosmological constant and pressure
\begin{equation}
P=-\frac{\Lambda }{8\pi },  \label{P}
\end{equation}
with the heat capacity
\begin{equation}
C_{Q}=\frac{T}{\left( \frac{\partial ^{2}M}{\partial S^{2}}\right) _{Q}}=%
\frac{T}{\left( \frac{\partial T}{\partial S}\right) _{Q}}.  \label{CQ}
\end{equation}

Using Eq. (\ref{TotalTT}), we will obtain the denominator of heat
capacity $\left( \frac{\partial T}{\partial S}\right) _{Q}$ in the
following form
\begin{eqnarray}
\left( \frac{\partial T}{\partial S}\right) _{Q} &=&\frac{\kappa d_{3}\left(
\mathcal{N}-2r_{+}^{2}\right) }{\pi d_{2}\mathcal{N}^{3}r_{+}^{d_{5}}}+\frac{%
\left( 3\mathcal{N}-2r_{+}^{2}\right) }{\pi d_{2}^{2}\mathcal{N}%
^{3}r_{+}^{d_{7}}}\left( \frac{4\beta ^{2}\left( \Upsilon _{+}-1\right) }{%
\Upsilon _{+}}-2\Lambda \right) -\frac{4h^{\prime }h^{\prime \prime }}{\pi
d_{2}^{2}\mathcal{N}^{2}r_{+}^{d_{8}}\Upsilon _{+}^{3}}- \nonumber \\
&&\frac{\alpha \kappa ^{2}d_{3}d_{4}d_{5}\left( \mathcal{N}%
+2r_{+}^{2}\right) }{\pi d_{2}\mathcal{N}^{3}r_{+}^{d_{3}}}-\frac{2m^{2}%
\mathcal{E}}{\pi d_{2}\mathcal{N}^{3}r_{+}^{d_{3}}}, \label{denom}
\end{eqnarray}%
in which $\mathcal{E}$ is
\begin{equation}
\mathcal{E}=d_{3}d_{4}\left[ d_{5}c^{4}c_{4}\left( r_{+}^{2}+\frac{\mathcal{N%
}}{2}\right) +c^{3}c_{3}r_{+}\right] +r_{+}^{2}\left[ d_{3}c^{2}c_{2}\left(
r_{+}^{2}-\frac{\mathcal{N}}{2}\right) +cc_{1}r_{+}\left( r_{+}^{2}-\mathcal{%
N}\right) \right].
\end{equation}

Now, by solving Eq. (\ref{denom}) with respect to pressure, a
relation for pressure is obtained
\begin{eqnarray}
P &=&\frac{m^{2}d_{2}}{8\pi r_{+}^{4}\left( r_{+}^{2}+6\alpha k\right) }%
\left\{ d_{3}d_{4}\left[ c_{4}c^{4}d_{5}\left(
\frac{3r_{+}^{2}}{2}+k\alpha \right) +c_{3}c^{3}r_{+}^{3}\right]
-r_{+}^{2}\left[ c_{2}c^{2}d_{3}\left( k\alpha
-\frac{r_{+}^{2}}{2}\right) +2\alpha c c_{1}k r_{+} \right]
\right\}  \nonumber \\
&&-\frac{\beta ^{2}\left( d_{3}r_{+}^{2}+2\alpha kd_{5}\right) \eta _{+}}{%
4\pi \left( r_{+}^{2}+6\alpha k\right) \sqrt{1+\eta
_{+}}}+\frac{kd_{2}\left[ \alpha k\left( 2 \alpha
d_{5}k+d_{9}r_{+}^{2}\right) +d_{3}r_{+}^{4}\right] }{16\pi
r_{+}^{4}\left( r_{+}^{2}+6\alpha k\right) }-\frac{\beta
^{2}}{4\pi }\left( 1-\frac{1}{\sqrt{1+\eta _{+}}}\right).
\end{eqnarray}

\begin{figure}[tbp]
$%
\begin{array}{ccc}
\epsfxsize=6.5cm \epsffile{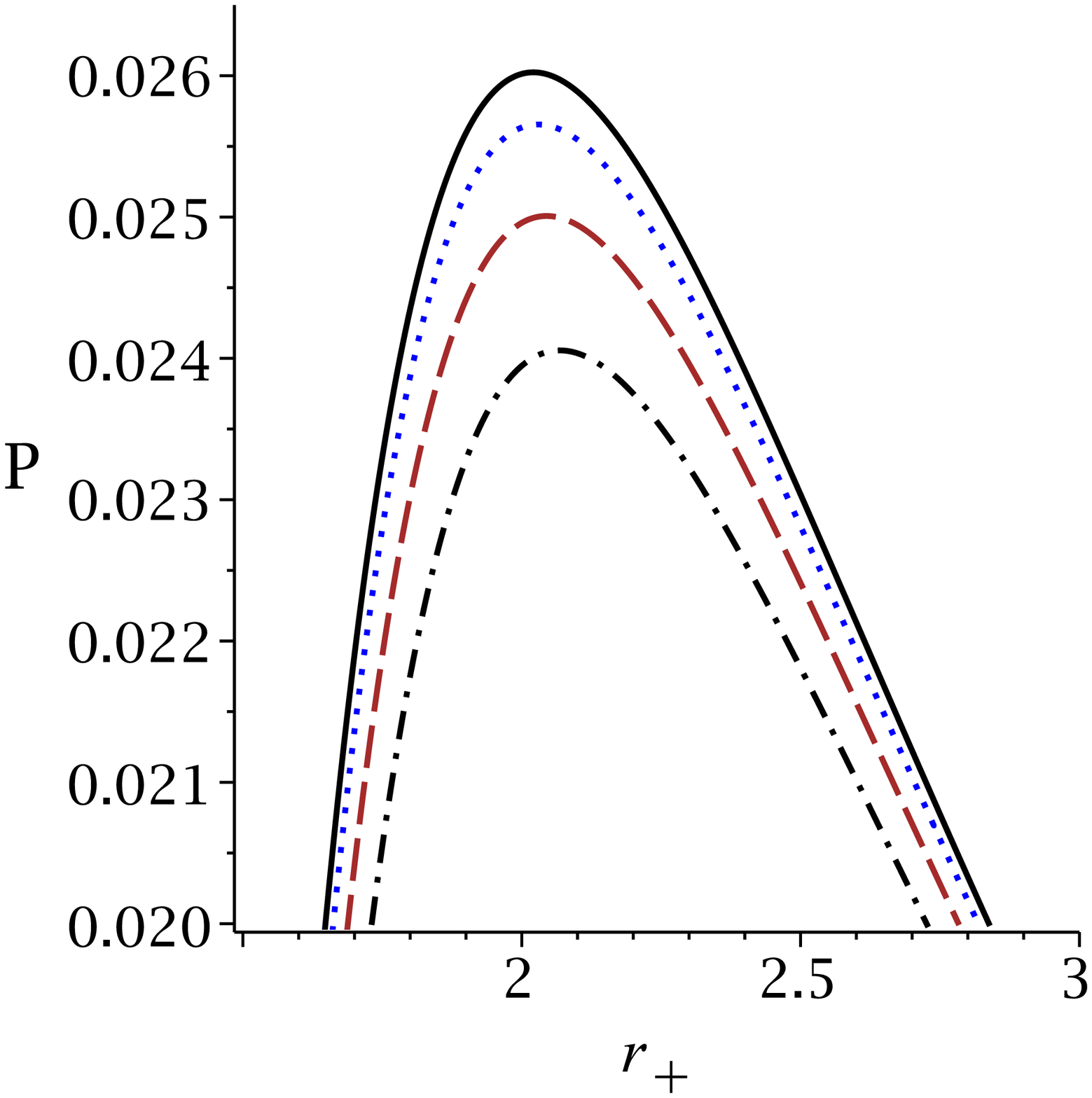} & \epsfxsize=6.5cm %
\epsffile{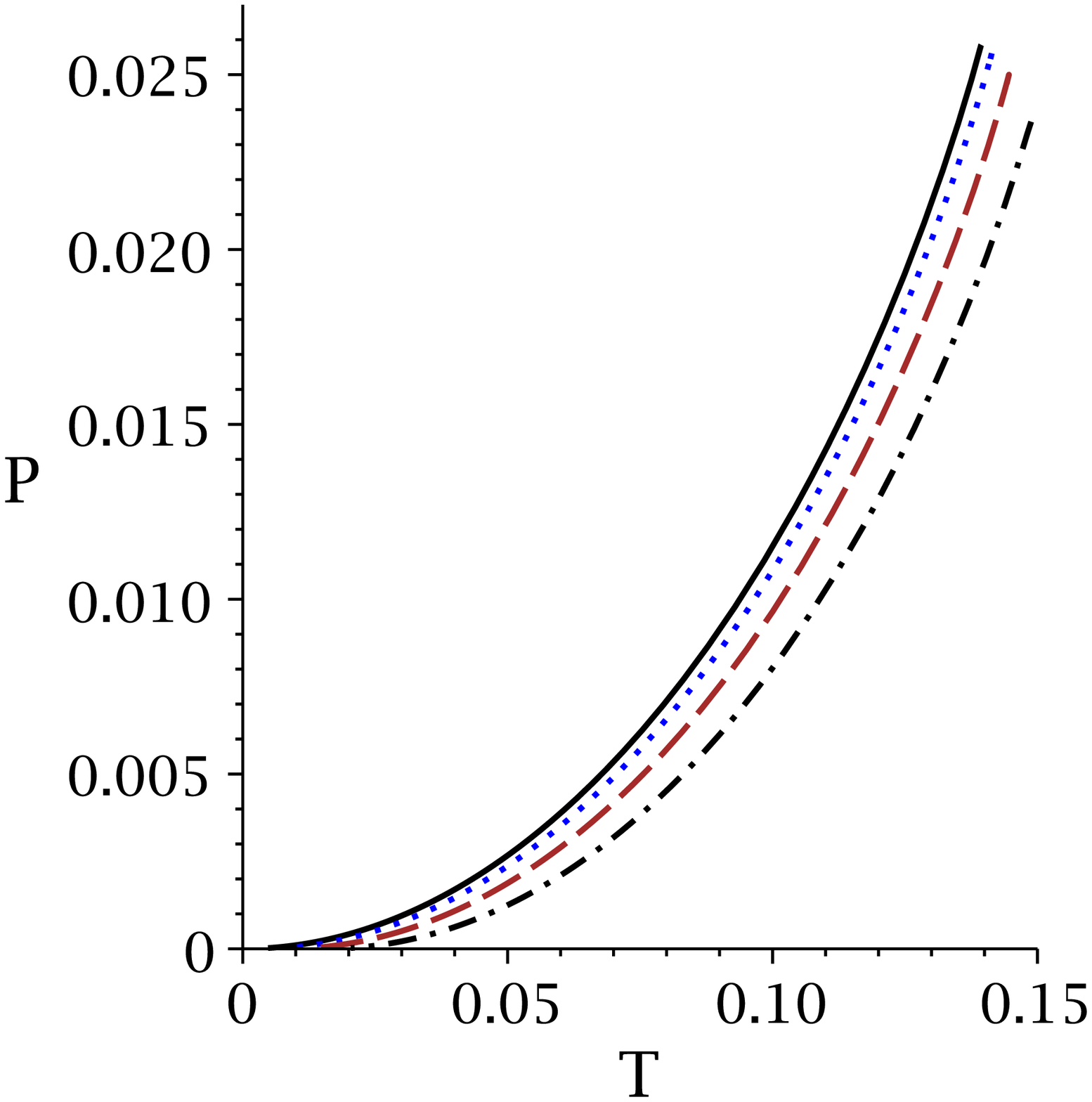} &
\end{array}
$%
\caption{$P$ versus $r_{+}$ (left panel) and $P$ versus $T$ (right panel)
for $q=1$, $c=c_{2}=c_{3}=c_{4}=0.2$, $c_{1}=2$, $\protect\beta=0.5$, $%
\protect\alpha=0.5$, $d=6$ and $k=1$; \newline Left panel: $m=0$
(continuous line), $m=0.3$ (dotted line), $m=0.5$ (dashed line)
and $m=0.7$ (dash-dotted line).} \label{Fig4}
\end{figure}

\begin{figure}[tbp]
$%
\begin{array}{ccc}
\epsfxsize=6.5cm \epsffile{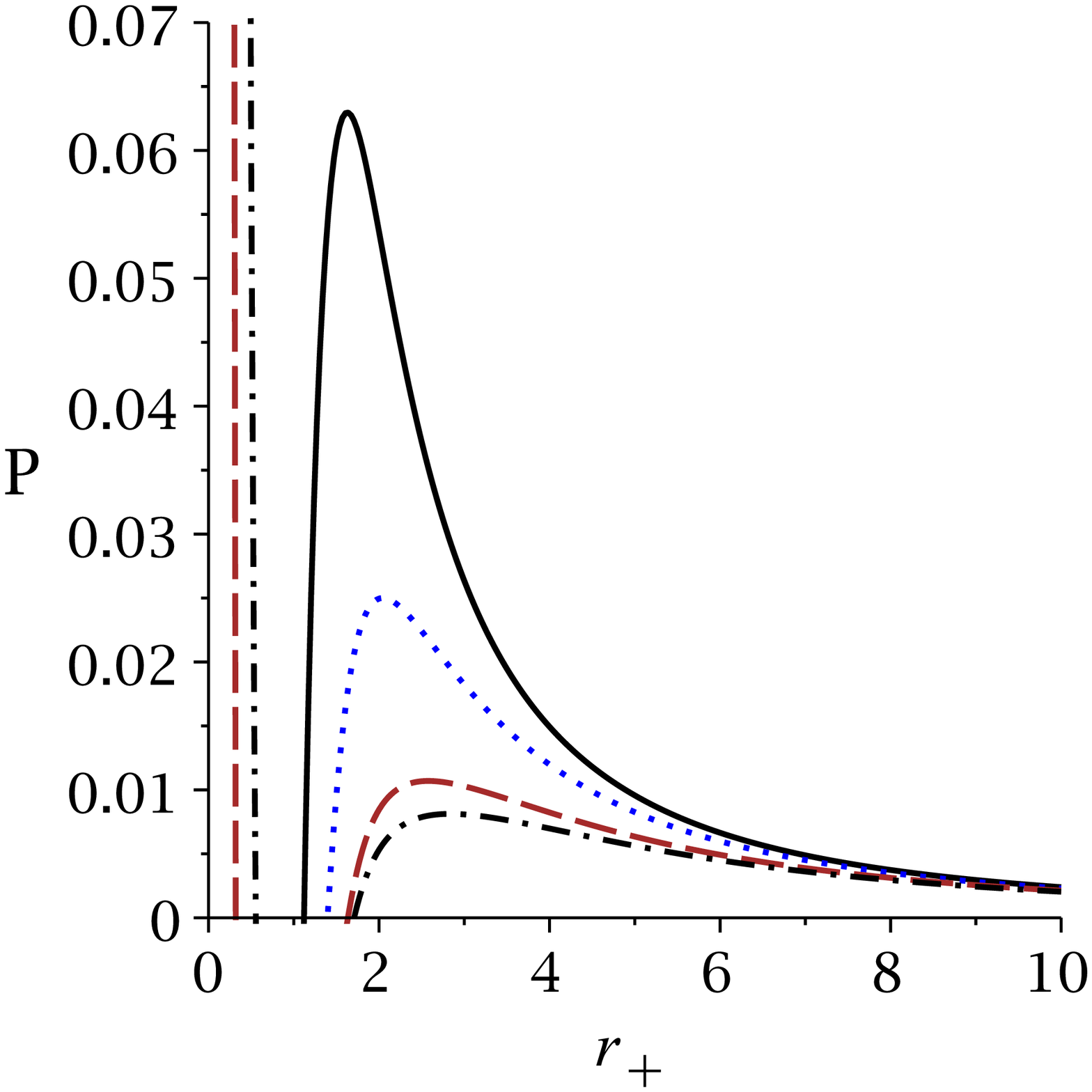} & \epsfxsize=6.5cm %
\epsffile{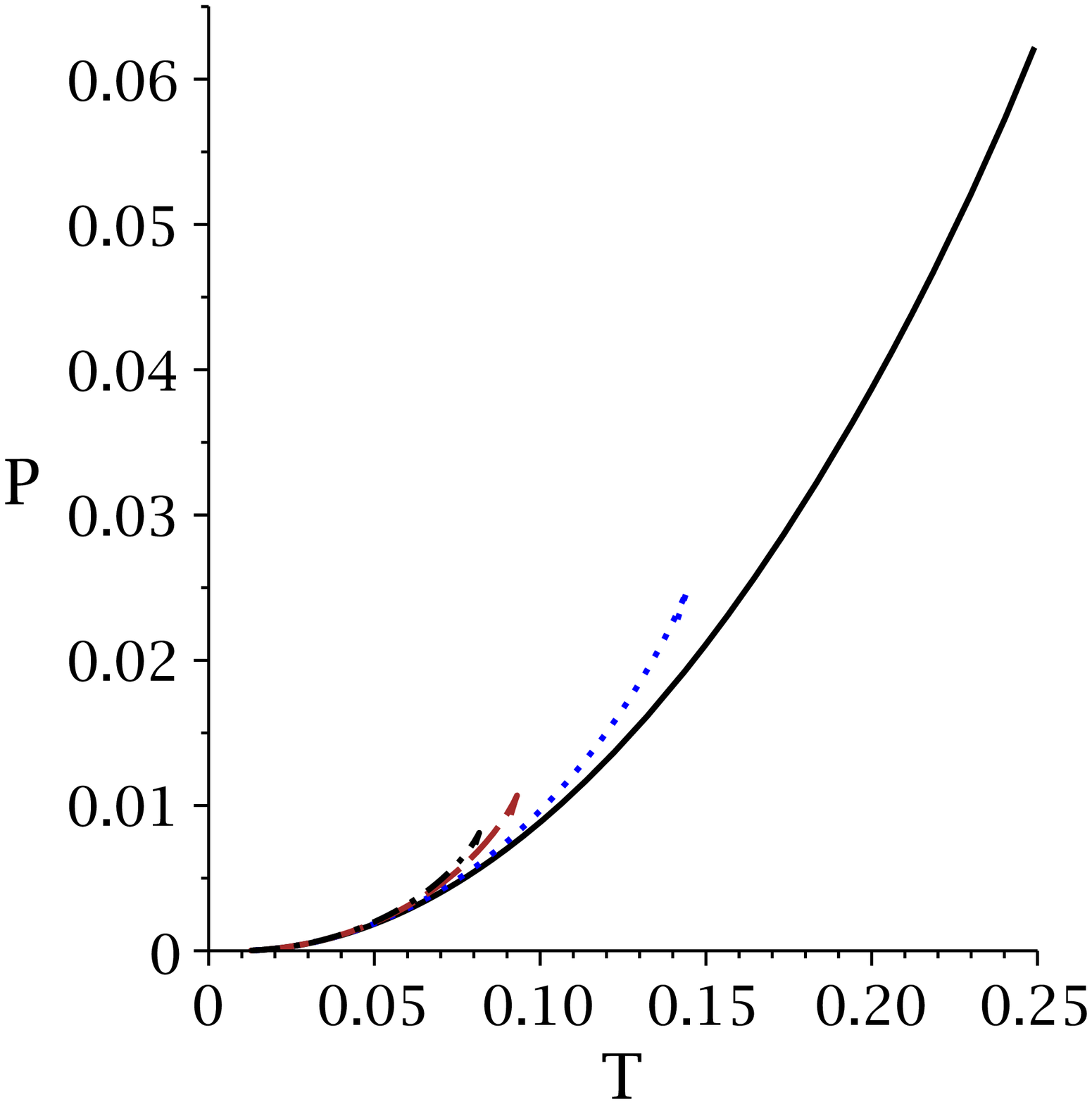} &
\end{array}
$%
\caption{$P$ versus $r_{+}$ (left panel) and $P$ versus $T$ (right panel)
for $q=1$, $c=c_{2}=c_{3}=c_{4}=0.2$, $c_{1}=2$ $m=0.5$, $\protect\beta=0.5$%
, $d=6$ and $k=1$; \newline
$\protect\alpha=0$ (continuous line), $\protect\alpha=0.5$ (dotted line), $%
\protect\alpha=1.5$ (dashed line) and $\protect\alpha=2$
(dash-dotted line).} \label{Fig5}
\end{figure}

\begin{figure}[tbp]
$%
\begin{array}{ccc}
\epsfxsize=5.5cm \epsffile{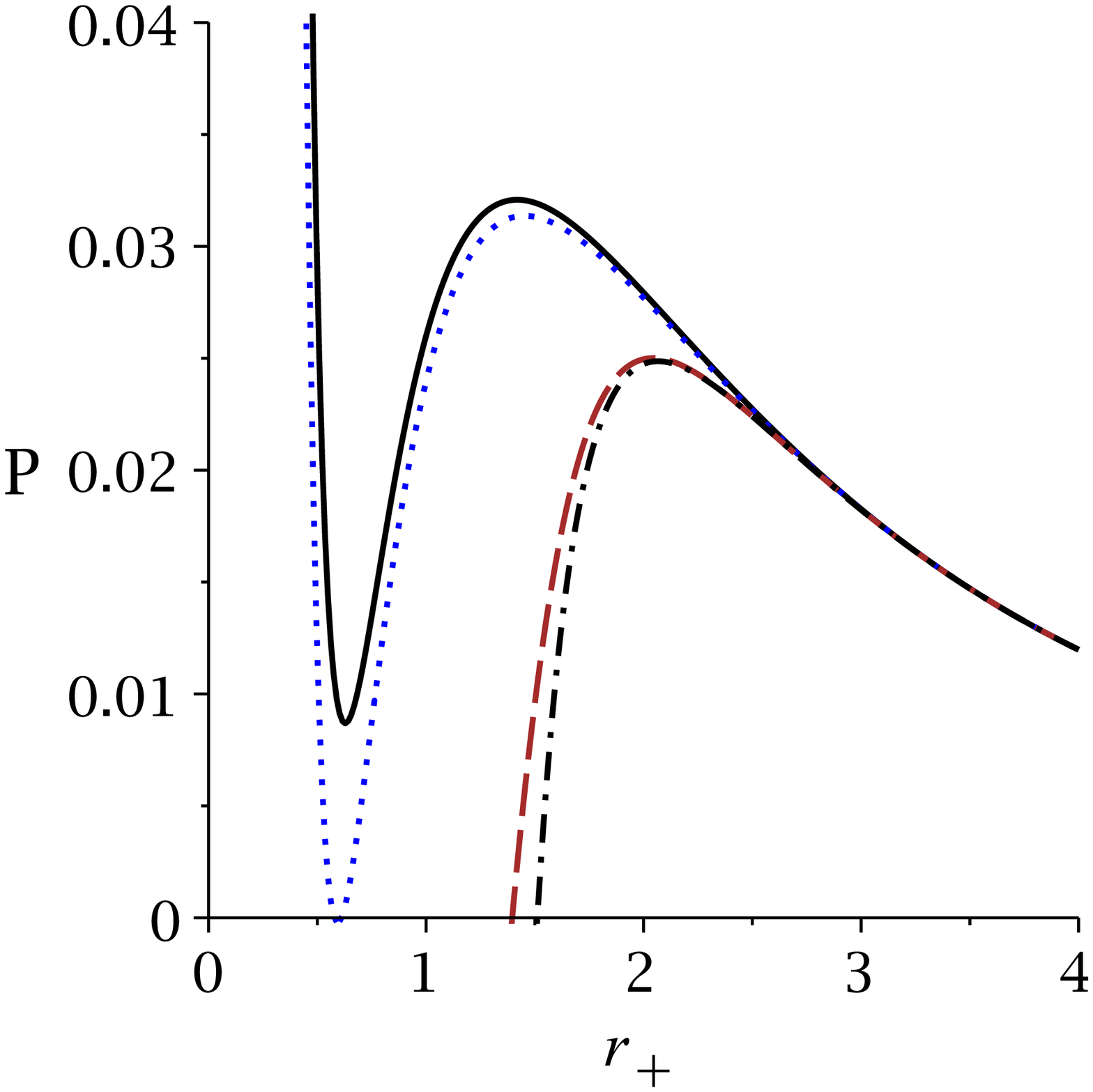} & \epsfxsize=5.5cm %
\epsffile{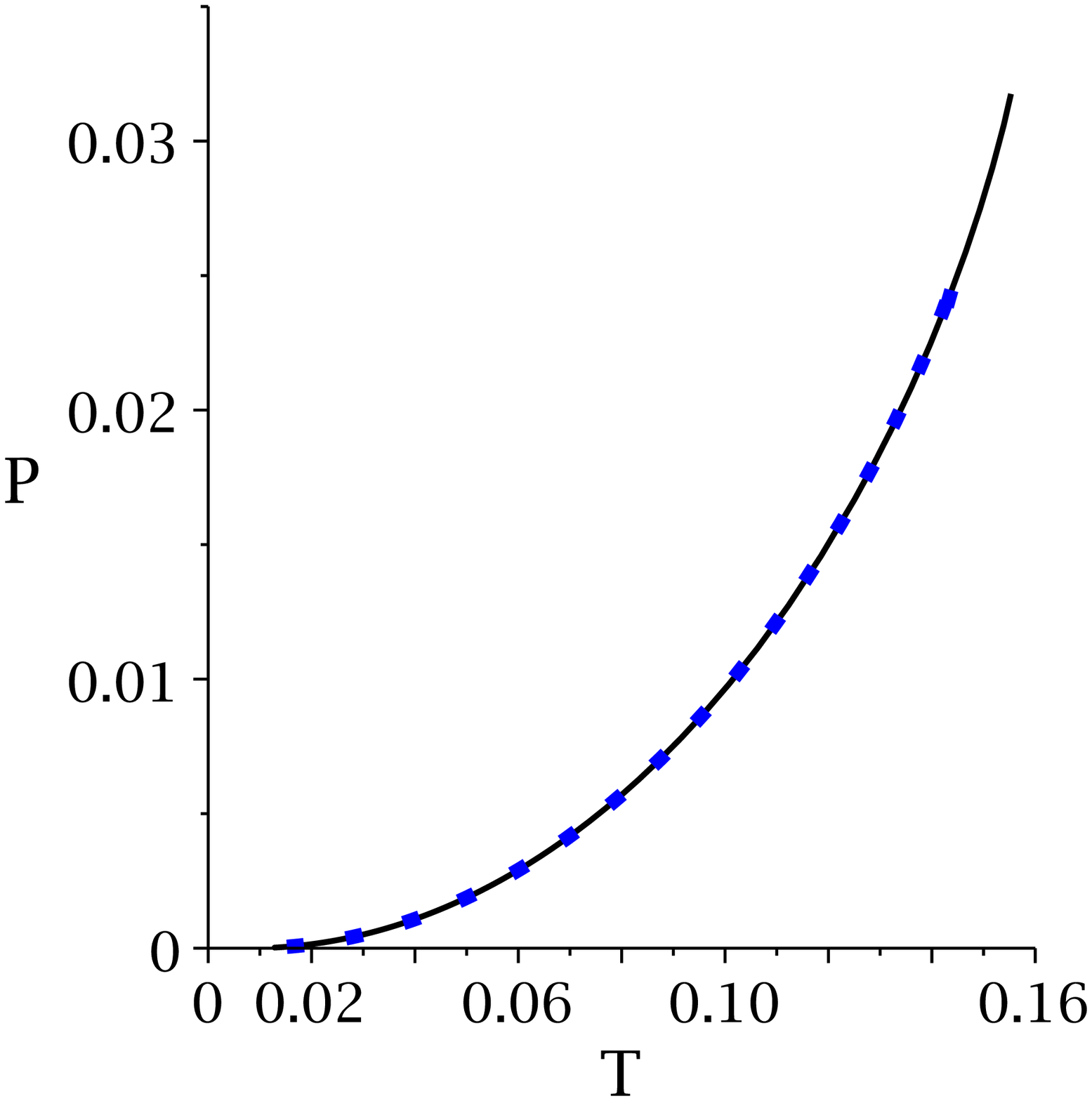} & \epsfxsize=5.5cm %
\epsffile{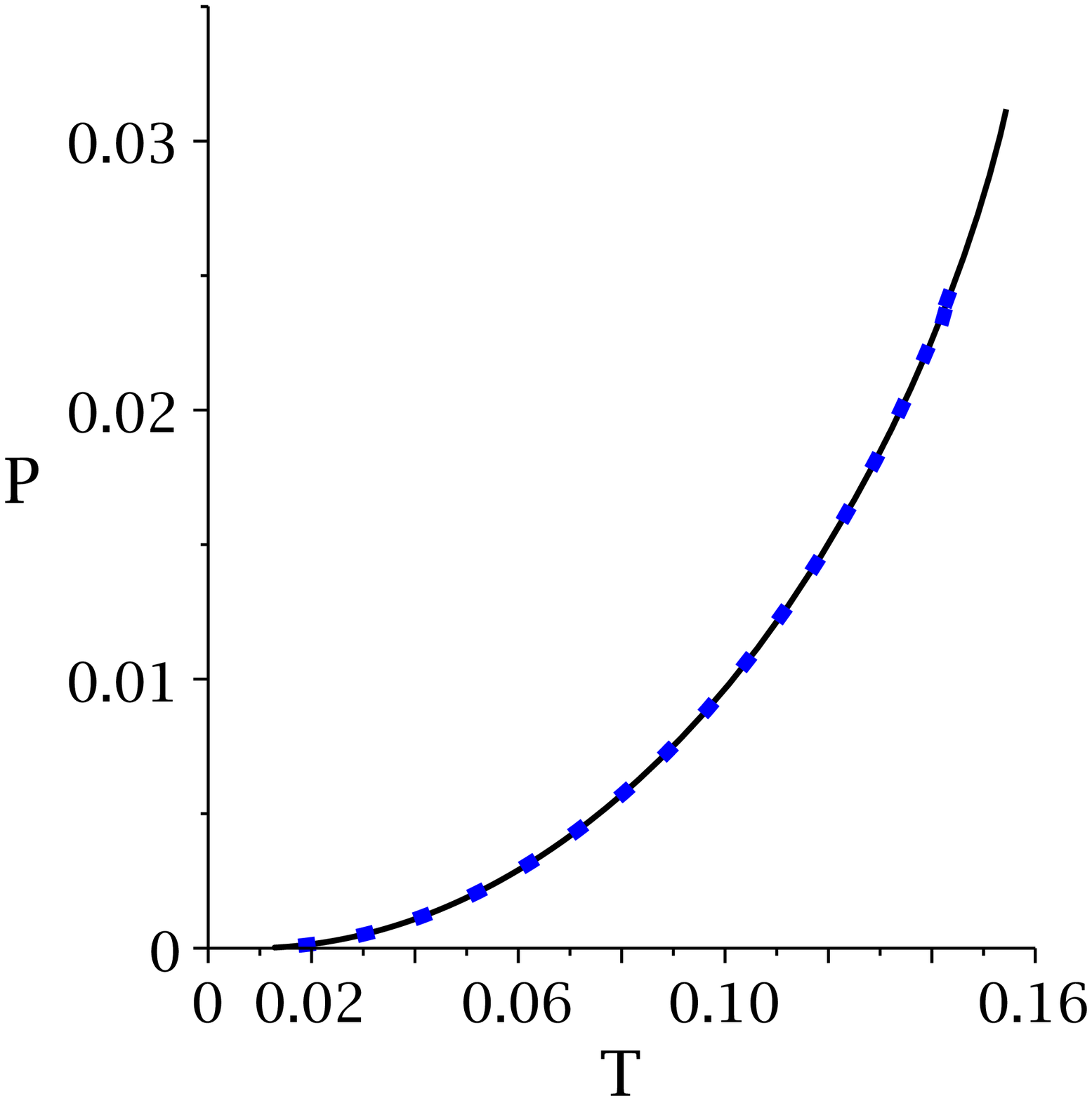}%
\end{array}
$%
\caption{$P$ versus $r_{+}$ (left panel) and $P$ versus $T$ (middle and
right panels) for $q=1$, $c=c_{2}=c_{3}=c_{4}=0.2$, $c_{1}=2$, $m=1$, $%
\protect\alpha=0.5$, $d=6$ and $k=1$; \newline Middle panel:
$\protect\beta=0.05$ (continuous line) and $\protect\beta=0.5$
(dotted line); Right panel: $\protect\beta=0.06$ (continuous line) and $%
\protect\beta=50$ (dotted line).} \label{Fig6}
\end{figure}

\begin{figure}[tbp]
$%
\begin{array}{ccc}
\epsfxsize=6.5cm \epsffile{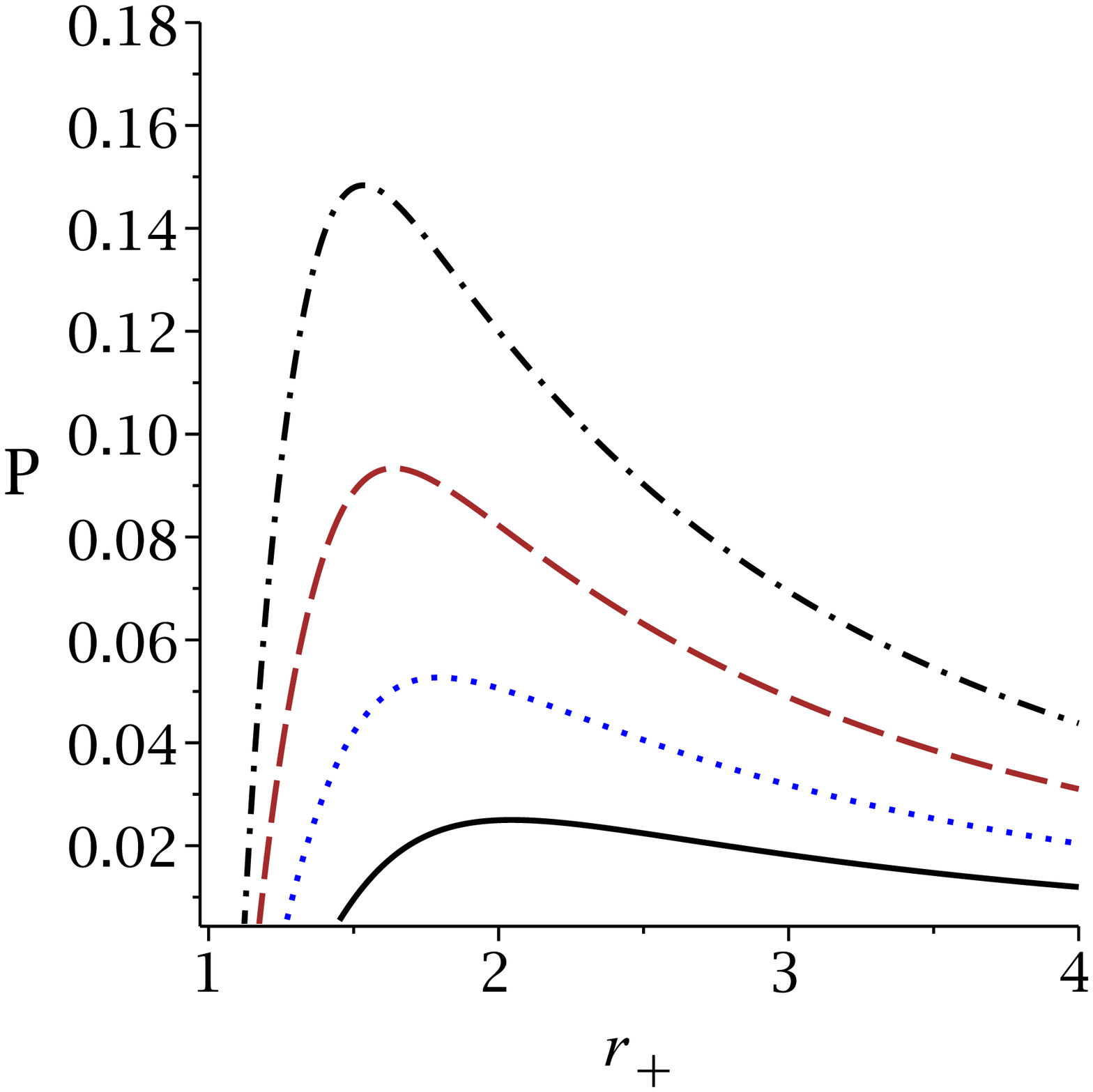} & \epsfxsize=6.5cm %
\epsffile{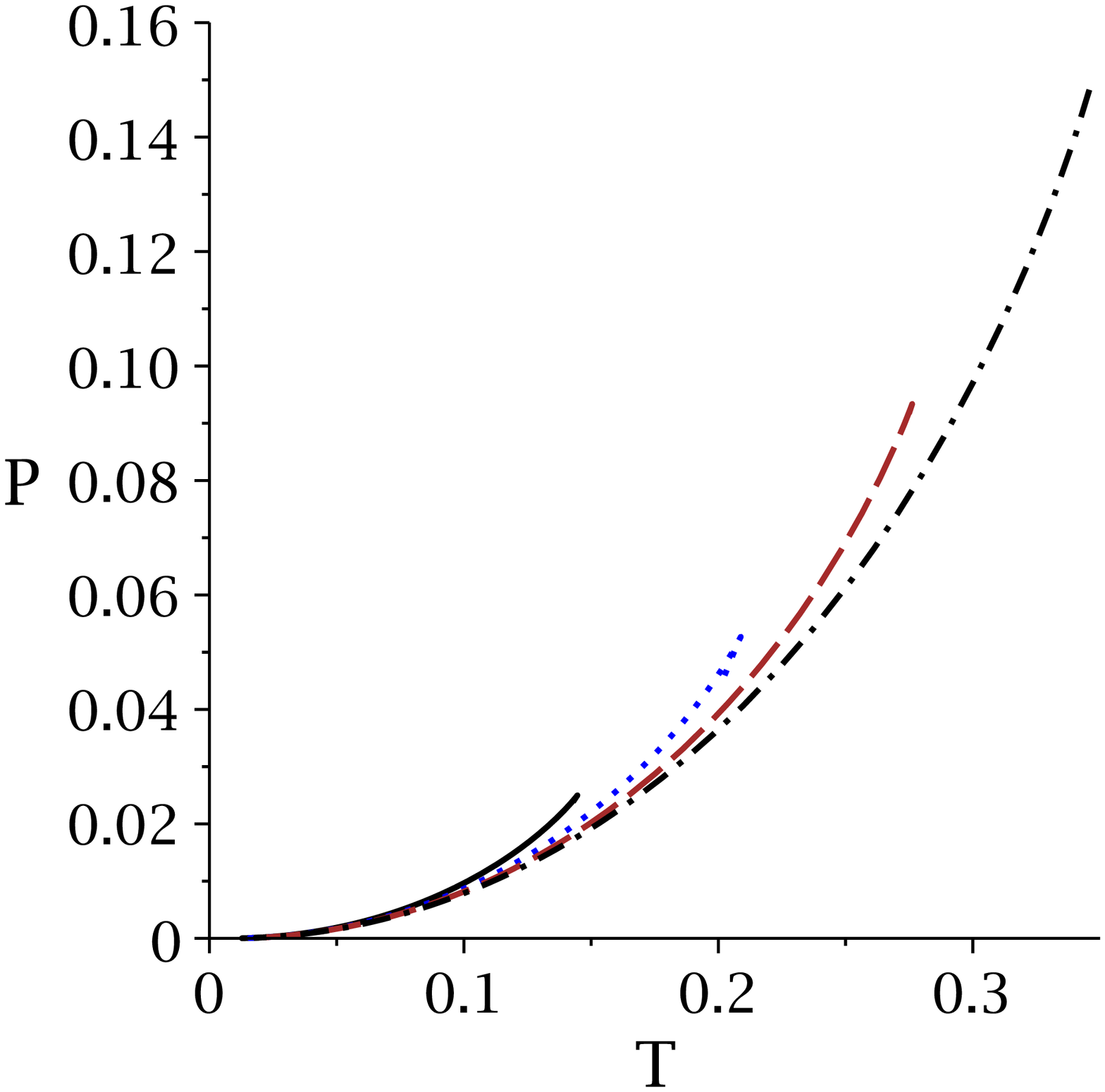} &
\end{array}
$%
\caption{$P$ versus $r_{+}$ (left panel) and $P$ versus $T$ (right panel)
for $q=1$, $c=c_{2}=c_{3}=c_{4}=0.2$, $c_{1}=2$, $m=1$, $\protect\alpha=0.5$%
, $\protect\beta=0.5$ and $k=1$; \newline $d=6$ (continuous line),
$d=7$ (dotted line), $d=8$ (dashed line) and $d=9$ (dash-dotted
line).} \label{Fig7}
\end{figure}

In order to study the critical behavior of these black holes, we should see
whether a maximum exists for this relation. To do so, we employ numerical
method. The results are presented in the following diagrams (Figs. \ref{Fig4}
-- \ref{Fig7}).

%%%%%%%%%%%%%%%%%%%%%%%%%%%%%%%%%%%%%%%%%%%%%%%%%%%%%%%%%%%%%%%

%%%%%%%%%%%%%%%%%%%%%%%%%%%%%%%%%%%%%%%%%%%%%%%%%%%%%%%%%%%%%%%
%%%%%%%%%%%%%%%%%%%%%%%%%%%%%%%%%%%%%%%%%%%%%%%%%%%%%%%%%%%%%%%

%%%%%%%%%%%%%%%%%%%%%%%%%%%%%%%%%%%%%%%%%%%%%%%%%%%%%%%%%%%%%%%
%%%%%%%%%%%%%%%%%%%%%%%%%%%%%%%%%%%%%%%%%%%%%%%%%%%%%%%%%%%%%%%

%%%%%%%%%%%%%%%%%%%%%%%%%%%%%%%%%%%%%%%%%%%%%%%%%%%%%%%%%%%%%%%
%%%%%%%%%%%%%%%%%%%%%%%%%%%%%%%%%%%%%%%%%%%%%%%%%%%%%%%%%%%%%%%

%%%%%%%%%%%%%%%%%%%%%%%%%%%%%%%%%%%%%%%%%%%%%%%%%%%%%%%%%%%%%%%

First of all, it is evident that due to existence of maximum,
these black holes enjoy a second order phase transition in their
phase space. The critical pressure is a decreasing function of the
massive, GB and BI parameters while their corresponding critical
horizon radius are increasing functions of them (left panels of
Figs. \ref{Fig4} - \ref{Fig6}). On the contrary, the critical
horizon radius is a decreasing function of the dimension while the
critical pressure is an increasing function of this parameter
(left panel of Fig. \ref{Fig7}).

Depending on choices of different parameters, one may come across
two interestingly different behavior for $P-r_{+}$ diagrams; I) in
one behavior, only one extremum exists for these diagrams (left
panels of Figs. \ref{Fig4} and \ref{Fig7}). II) in the other one,
one minimum and one maximum exist (left panels of Figs. \ref{Fig5}
and \ref{Fig6}).

Considering mentioned concept for this method, only in maximum a
second order phase transition exists. Therefore, we have a second
order phase transition for both cases. On the other hand, for
second behavior, for critical pressure, two horizon radii exist
(due to formation of tail). This indicates that another branch for
critical behavior exists. This critical behavior is not a second
order phase transition but rather another kind (considering the
concept of maximum). Interestingly, the second case of behavior
for $P-r_{+}$ observed for small values of nonlinearity parameter
and large values of GB parameter. In other words, for small values
of BI parameter and large values of GB parameter, existence of
extra branch in
phase diagrams of these black holes is evident (left panels of Figs. \ref%
{Fig5} and \ref{Fig6}).

Existence of such behavior points out that another branch for
phase diagrams exists for these black holes which is absent in
other black holes. Such behavior is precisely due to existence of
massive gravity. This means that by considering a massive theory
of gravity for these black holes, another type of phase transition
takes place. This emphasizes on the role and effects of massive
gravity in thermodynamical behavior of these black holes.

In order to complete our study here, we will plot coexistence curves for
variation of different parameters as well (right panels of Figs. \ref{Fig4} - %
\ref{Fig7}). The coexistence curves are representing small/larger
black holes with similar pressure and temperature. The critical
point is located at the end of this line which indicating after
this point, phase transition does not take place. Evidently, the
critical temperature is an increasing function of the massive
gravity (right panel of Fig. \ref{Fig4}) and dimensionality (right
panel of Fig. \ref{Fig7}) while it is a decreasing function of the
GB (right panel of Fig. \ref{Fig5}) and BI parameters (right panel
of Fig. \ref{Fig6}). Here, in these phase diagrams, we see that
the presence of other phase transition is not observed. This
indicates that our earlier interpretation is right. In other
words, the branch for phase transition which was observed in
$P-r_{+}$ diagram is not a second order phase transition. Also, we
should point out that plotted diagrams indicate that no reentering
of phase transition takes place for these black holes.

\section{Check of Maxwell equal area law for both $T-S$ and $P-V$ graphs
\label{MEAL}}

The expressions of Hawking temperature and the entropy are listed
in Eqs. (\ref{TotalTT}) and (\ref{TotalS}) respectively. For
$T-r_+$ graph, the possible critical point can be determined
through
\begin{eqnarray}
\left(\frac{\partial T}{\partial r_+}\right)_{q=q_c, r=r_{+c}}&=0,
\label{701} \\
\left(\frac{\partial^2 T}{\partial r_+^2}\right)_{q=q_c, r=r_{+c}}&=0.
\label{702}
\end{eqnarray}

For $T-S$ graph, the possible critical point can be determined through
\begin{eqnarray}
\left(\frac{\partial T}{\partial S}\right) _{q=q_c, S=S_c}&=0,  \label{703}
\\
\left(\frac{\partial^2 T}{\partial S^2}\right) _{q=q_c, S=S_c}&=0.
\label{704}
\end{eqnarray}

Eqs. (\ref{703}) and (\ref{704}) are related to Eqs. (\ref{701}) and (\ref%
{702}) by
\begin{eqnarray}
\left(\frac{\partial T}{\partial S}\right)&=& \left(\frac{\partial T}{%
\partial r_+}\right)/\left(\frac{\partial S}{\partial r_+}\right),
\label{705} \\
\left(\frac{\partial^2 T}{\partial S^2}\right)&=& \left(\frac{
 \partial \left(\frac{\partial
T}{\partial S}\right)    }{\partial r_+}\right)/\left(\frac{\partial S}{\partial r_+}%
\right)=\left(\frac{\left(\frac{\partial^2 T}{\partial r_+^2}\right)\left(%
\frac{\partial S}{\partial r_+}\right)-\left(\frac{\partial T}{\partial r_+}%
\right)\left(\frac{\partial^2 S}{\partial r_+^2}\right)}{\left(\frac{%
\partial S}{\partial r_+}\right)^2}\right)/\left(\frac{\partial S}{\partial
r_+}\right).  \label{706}
\end{eqnarray}%

Considering the above two relations and the fact that $\left(\frac{\partial S%
}{\partial r_+}\right)>0$, it is not difficult to conclude that
the critical point conditions for $T-r_+$ and $T-S$ graphs are
equivalent to each other.

To probe the effect of massive gravity on critical quantities of
$T-S$ graph, we fix other parameters and let $m$ vary from $0$ to
$0.3$. The results are listed in Table \ref{tb701}. One can see
clearly that for the cases $m=0$ and $m=0.1$, there are two
critical points while there is only one for the cases $m=0.2$ and
$m=0.3$. Then, we let $\alpha$ vary and keep other parameters
fixed to investigate the effect of GB gravity. The results are
listed in Table \ref{tb702}. Lastly, we let $\beta$ vary and keep
other parameters fixed to study the effect of BI theory. The
results are listed in Table \ref{tb703}.
\begin{table}[!h]
\tabcolsep 0pt \caption{Effect of $m$ on critical quantities of
$T-r_+$ graph for $\alpha=0.5, \beta=0.5, c=c_1=c_2=2, c_3=0.2,
c_4=-0.2, d=6, \Lambda=-0.1$} \vspace*{-12pt}
\begin{center}
\def\temptablewidth{1\textwidth}
{\rule{\temptablewidth}{2pt}}
\begin{tabular*}{\temptablewidth}{@{\extracolsep{\fill}}ccccccc}
$m$ & $q_{c1}$ & $r_{c1}$ &$T_{c1}$ & $q_{c2}$ & $r_{c2}$
&$T_{c2}$ \\   \hline
    0    & 19.18887563    & 5.35344750      &      0.05519929  & 0.53171174&  1.27585252 &0.06336630\\
    0.1 & 22.42316953     & 5.58508070       &      0.06071315 & 0.30110234& 1.54894737&0.06801836 \\
 0.2   & 35.70639390    & 6.28495138       &       0.07690367  & - & - &-\\
  0.3   &70.40090638   & 7.39745021        &        0.10284671  &- & - &-
          \end{tabular*}
       {\rule{\temptablewidth}{2pt}}
       \end{center}
       \label{tb701}
       \end{table}

\begin{table}[!h]
\tabcolsep 0pt \caption{Effect of $\alpha$ on critical quantities
of $T-r_+$ graph for $m=0.1, \beta=0.5, c=c_1=c_2=2, c_3=0.2,
c_4=-0.2, d=6, \Lambda=-0.1$} \vspace*{-12pt}
\begin{center}
\def\temptablewidth{1\textwidth}
{\rule{\temptablewidth}{2pt}}
\begin{tabular*}{\temptablewidth}{@{\extracolsep{\fill}}ccccccc}
$\alpha$ & $q_{c1}$ & $r_{c1}$ &$T_{c1}$ & $q_{c2}$ & $r_{c2}$
&$T_{c2}$ \\   \hline
    0    & 67.53285621  & 6.97544318     &     0.06666372  & -&  - &-\\
    0.3 & 38.80127104    & 6.21579271       &     0.06323488 & 0.16299321& 1.05596356&0.08587341 \\
 0.5 & 22.42316953     & 5.58508070       &      0.06071315 & 0.30110234& 1.54894737&0.06801836 \\
  0.7  &7.91485816   & 4.69335274       &        0.05786188  &- & - &-
          \end{tabular*}
       {\rule{\temptablewidth}{2pt}}
       \end{center}
       \label{tb702}
       \end{table}

\begin{table}[!h]
\tabcolsep 0pt \caption{Effect of $\beta$ on critical quantities
of $T-r_+$ graph for $\alpha=0.5, \beta=0.5, c=c_1=c_2=2, c_3=0.2,
c_4=-0.2, d=6, \Lambda=-0.1$} \vspace*{-12pt}
\begin{center}
\def\temptablewidth{1\textwidth}
{\rule{\temptablewidth}{2pt}}
\begin{tabular*}{\temptablewidth}{@{\extracolsep{\fill}}ccccccc}
$\beta$ & $q_{c1}$ & $r_{c1}$ &$T_{c1}$ & $q_{c2}$ & $r_{c2}$
&$T_{c2}$ \\   \hline
    0.1    & 24.63144852    & 5.42252911      &      0.06064608 & 0.43456369&  1.66031363 &0.06791006 \\
  0.5 & 22.42316953     & 5.58508070       &      0.06071315 & 0.30110234& 1.54894737&0.06801836 \\
 1   & 22.36488018    & 5.58910963      &      0.06071483  & 0.29659468 & 1.53439231 &0.06802347 \\
  2   &22.35039262  &5.59010866        &       0.06071524  &0.29542777 & 1.52977250 &0.06802485
          \end{tabular*}
       {\rule{\temptablewidth}{2pt}}
       \end{center}
       \label{tb703}
       \end{table}

%%%%%%%%%%%%%%%%%%%%%%%%%%%%%%%%%%%%%%%%%%%%%%%%%%%%%%%%%%%%%%%%%%%%%%%%%%%%%
\begin{figure*}
\centerline{\subfigure[]{\label{701a}
\includegraphics[width=8cm,height=6cm]{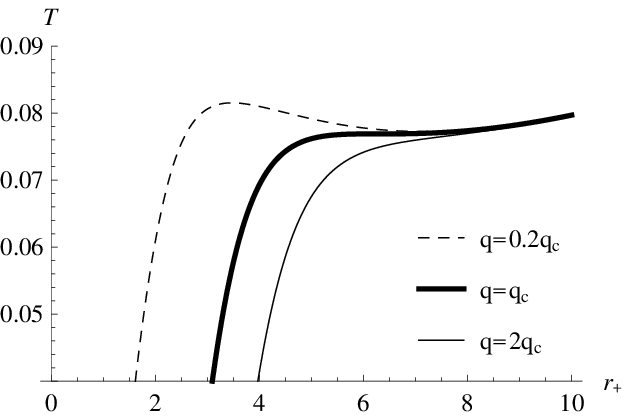}}
\subfigure[]{\label{701b}
\includegraphics[width=8cm,height=6cm]{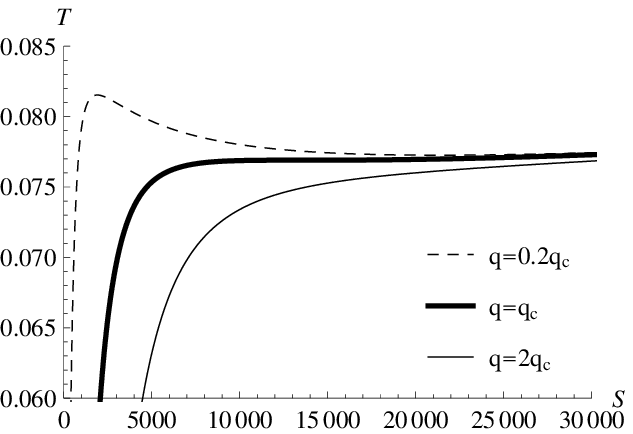}}}
\caption{(a) $T$ vs. $r_+$ for $m=0.2, \alpha=0.5, \beta=0.5,
c=c_1=c_2=2, c_3=0.2, c_4=-0.2, d=6, \Lambda=-0.1$ (b) $T$ vs. $S$
for $m=0.2, \alpha=0.5, \beta=0.5, c=c_1=c_2=2, c_3=0.2, c_4=-0.2,
d=6, \Lambda=-0.1$} \label{fg701}
\end{figure*}
%%%%%%%%%%%%%%%%%%%%%%%%%%%%%%%%%%%%%%%%%%%%%%%%%%%%%%%%%%%%%%%%%%%%%%%%%%%%%%%%
%%%%%%%%%%%%%%%%%%%%%%%%%%%%%%%%%%%%%%%%%%%%%%%%%%%%%%%%%%%%%%%%%%%%%%%%%%%%%%%%

To gain an intuitive understanding of the Van der Waals like
behavior, we plot both $T-r_+$ and $T-S$ graphs for the case
$m=0.2, \alpha=0.5, \beta=0.5, c=c_1=c_2=2, c_3=0.2, c_4=-0.2,
d=6, \Lambda=-0.1$. From Fig. \ref{fg701}, one can see clearly
that both the graphs can be divided into three branches. The
medium radius branch is unstable while both the large radius
branch and the small radius branch are stable. The unstable branch
in $T-S$ curve can be removed with a bar vertical to the
temperature axis $T=T_*$ as the approach in Ref. \cite{Spallucci}.
The possible Maxwell equal area laws for $T-S$ and $T-r_+$ graphs
read
\begin{eqnarray}
T_*(S_3-S_1)&=&\int^{S_3}_{S_1}TdS,  \label{707} \\
T_*(r_3-r_1)&=&\int^{r_3}_{r_1}Tdr_+.  \label{708}
\end{eqnarray}

Note that $S_1$, $S_2$, $S_3$ denote the three values of entropy
from small to large corresponding to $T=T_*$ while $r_1$, $r_2$,
$r_3$ denote the three values of $r_+$ from small to large
corresponding to $T=T_*$.

To determine $T_*$, one should first study the behavior of free
energy ($F$), which can be obtained as
\begin{eqnarray}
F&=&M-TS=-\frac{V_{d_2}r_{+}^{d_5}(r_{+}^{2}+2d_3d_2\alpha\kappa)}
{16\pi
(r_+^2+2d_4d_3\alpha\kappa)}\Big\{m^2[cr_+^2(cc_2d_3+c_1r_+)
+c^3d_4d_3(cc_4d_5+c_3r_+)]+d_3\kappa(r_+^2+d_5d_4\alpha\kappa)  \notag \\
&\;&+\frac{2r_{+}^{4}(2\beta^2-\Lambda-2\beta^2\sqrt{1+\eta_+})} {d_2}\Big\}+%
\frac{V_{d_2}}{16\pi}m^2r_+^{d_5}[cr_+^2(cc_2d_2+c_1r_+)
+c^3d_3d_2(cc_4d_4+c_3r_+)]  \notag \\
&\;&-\frac{V_{d_2}r_+^{d_1}(\Lambda-2\beta^2+2\beta^2 \sqrt{1+\eta_+})}{8\pi
d_1}+\frac{V_{d_2}d_2}{16\pi}\left(r_+^{d_3}\kappa+d_3
d_4r_+^{d_5}\alpha\kappa^2+\frac{2d_2q^2r_{+}^{-d_3} \mathcal{H}_+}{d_1}%
\right).  \label{709}
\end{eqnarray}

We plot the free energy for the case $q=0.2q_c, m=0.2, \alpha=0.5,
\beta=0.5, c=c_1=c_2=2, c_3=0.2, c_4=-0.2, d=6, \Lambda=-0.1$ in Fig. \ref%
{fg702}, where we can find the swallow tail characteristic of first order
phase transition.
%%%%%%%%%%%%%%%%%%%%%%%%%%%%%%%%%%%%%%%%%%%%%%%%%%%%%%%%%%%%%%%%%%%%%%%%%%%%%
\begin{figure*}[tbp]
\centerline{\subfigure[]{\label{702a}
\includegraphics[width=8cm,height=6cm]{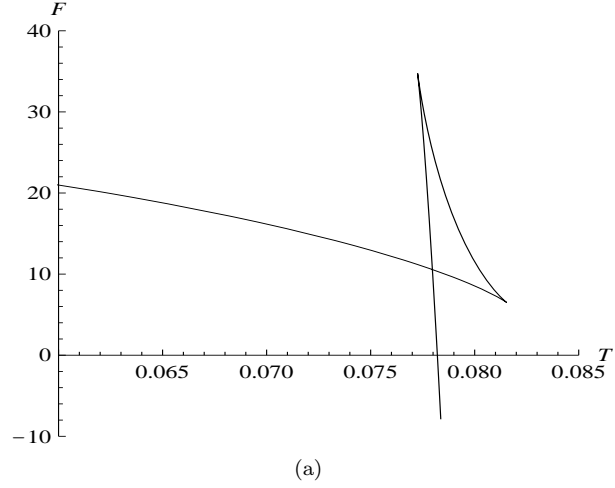}}}
\caption{$F$ vs. $T$ for $q=0.2q_c, m=0.2, \protect\alpha=0.5, \protect\beta%
=0.5, c=c_1=c_2=2, c_3=0.2, c_4=-0.2, d=6, \Lambda=-0.1$ }
\label{fg702}
\end{figure*}
%%%%%%%%%%%%%%%%%%%%%%%%%%%%%%%%%%%%%%%%%%%%%%%%%%%%%%%%%%%%%%%%%%%%%%%%%%%%%%%%
Numerical check of Maxwell equal area law for the cases $q=0.2q_c, 0.4q_c,$ $%
0.6q_c, 0.8q_c$ are carried out in both the $T-r_+$ and $T-S$
graphs. The first order phase transition temperature $T_*$ is
obtained through the intersection point of two branches in the
free energy curve. As shown in Table. \ref{tb705} and \ref{tb706},
the relative errors are very small and the Maxwell equal area law
holds for not only $T-r_+$ curves, but also $T-S$ curves.

\begin{table}[!h]
\tabcolsep 0pt \caption{Numerical check of Maxwell equal area law
for $T-r_+$ graph for $m=0.2, \alpha=0.5, \beta=0.5, c=c_1=c_2=2,
c_3=0.2, c_4=-0.2, d=6, \Lambda=-0.1$ } \vspace*{-12pt}
\begin{center}
\def\temptablewidth{1\textwidth}
{\rule{\temptablewidth}{2pt}}
\begin{tabular*}{\temptablewidth}{@{\extracolsep{\fill}}cccccccc}
$q$ & $T_*$ & $r_1$ &$r_2$ &$r_3$ &$T_*(r_3-r_1)$ &
$\int^{r_3}_{r_1}Tdr_+$ & relative error \\   \hline
    $0.2q_c$    & 0.077968093     & 2.668179599      &      5.831521024& 8.610961493 & 0.463347371 & 0.468423238 & $1.08361\times10^{-2}$ \\
     $0.4q_c$  & 0.077709445    & 3.460484711     &     5.961743318 & 8.320522932  & 0.377670873 & 0.379524375 & $4.88375\times10^{-3}$ \\
 $0.6q_c$     &  0.077443230     &  4.150221293       & 6.082516075       &7.965173990 & 0.295442259 & 0.295984384 & $1.83160\times10^{-3}$ \\
   $0.8q_c$    & 0.077174217    & 4.875672340       &      6.189311151 & 7.492034394 & 0.201915693 & 0.201996209 & $3.98602\times10^{-4}$
       \end{tabular*}
       {\rule{\temptablewidth}{2pt}}
       \end{center}
       \label{tb705}
       \end{table}

\begin{table}[!h]
\tabcolsep 0pt \caption{Numerical check of Maxwell equal area law
for $T-S$ graph for $m=0.2, \alpha=0.5, \beta=0.5, c=c_1=c_2=2,
c_3=0.2, c_4=-0.2, d=6, \Lambda=-0.1$ } \vspace*{-12pt}
\begin{center}
\def\temptablewidth{1\textwidth}
{\rule{\temptablewidth}{2pt}}
\begin{tabular*}{\temptablewidth}{@{\extracolsep{\fill}}cccccccc}
$q$ & $T_*$ & $S_1$ &$S_2$ &$S_3$ &$T_*(S_3-S_1)$ &
$\int^{S_3}_{S_1}TdS$ & relative error \\   \hline
    $0.2q_c$    & 0.077968093     & 895.587288070     &     10294.201627482 & 42030.087706891 & 3207.178554163 & 3207.178545951 & $2.56051\times10^{-9}$ \\
     $0.4q_c$  & 0.077709445    & 1889.035616718    &     11118.238437108 & 37002.585804846  & 2728.654497099 & 2728.654509847 & $4.67190\times10^{-9}$ \\
 $0.6q_c$     &  0.077443230     &  3312.040653402       & 11927.365360006       &31493.701542953 & 2182.478846051 & 2182.478875402 & $1.34485\times10^{-8}$ \\
   $0.8q_c$    & 0.077174217    & 5595.294205719      &     12680.211012687 & 25162.285938926 & 1510.067266056 & 1510.067258972 & $4,69118\times10^{-9}$
       \end{tabular*}
       {\rule{\temptablewidth}{2pt}}
       \end{center}
       \label{tb706}
       \end{table}

The possible Maxwell equal area laws for $P-r_+$ and $P-V$ graphs
read
\begin{eqnarray}
P_*(r_3-r_1)&=&\int^{r_3}_{r_1}Pdr_+,  \label{710} \\
P_*(V_3-V_1)&=&\int^{V_3}_{V_1}PdV.  \label{711}
\end{eqnarray}

Here, $r_1$, $r_2$ and $r_3$ denote the three values of $r_+$ from small to
large corresponding to $P=P_*$ in $P-r_+$ graph while $V_1$, $V_2$ and $V_3$
denote the three values of $V$ from small to large corresponding to $P=P_*$
in $P-V$ graph. Note that thermodynamic volume $V$ is defined in the
extended phase space as $V=\left(\frac{\partial M}{\partial P}\right)_{S,Q}$%
. For the cases $P_*=0.5P_c, 0.6P_c,$ $0.7P_c, 0.8P_c$, we use the
technique of Gibbs free energy to determine the corresponding
$T_*$, which is shown in the first column of Tables. \ref{tb707}
and \ref{tb708}. Since the mass of black hole should be
interpreted as enthalpy in the extended phase space, the
definition of Gibbs free energy reads
\begin{equation}
G=H-TS=M-TS.  \label{712}
\end{equation}

\begin{figure*}[tbp]
\centerline{\subfigure[]{\label{703a}
\includegraphics[width=8cm,height=6cm]{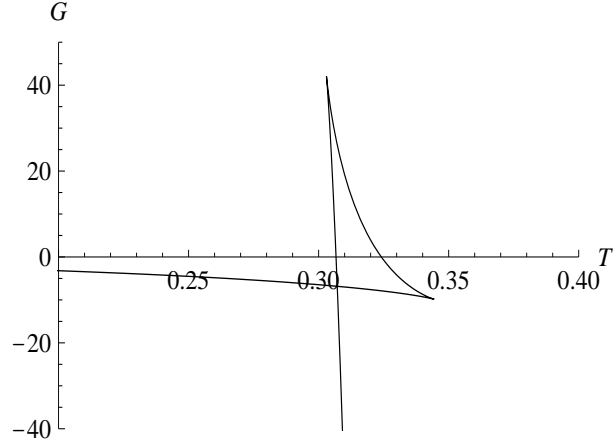}}}
\caption{$G$ vs. $T$ for $P_*=0.5P_c, m=0.5, \protect\alpha=0.8/6, \protect%
\beta=0.5, c=c_1=c_2=2, c_3=0.2, c_4=-0.2, d=6, q=1$ }
\label{fg703}
\end{figure*}

We plot the Gibbs free energy for the case $P_*=0.5P_c, m=0.5, \alpha=0.8/6,
\beta=0.5, c=c_1=c_2=2, c_3=0.2, c_4=-0.2, d=6, q=1$ in Fig. \ref{fg703}.
The classical swallow tail behavior can also be found. We further calculate
both the left hand side and right hand side of Eqs. (\ref{710}) and (\ref%
{711}). As shown in Tables. \ref{tb707} and \ref{tb708}, the relative errors
for $P-r_+$ graph are very large while those for $P-V$ graph are amazingly
small, leading to the conclusion that the Maxwell equal area law holds for $%
P-V$ graph while it fails for $P-r_+$ graph. Our numerical results here for
the GB-BI-massive black holes further backup the findings in former
researches \cite{Shaowen2,lanshanquan}.
%%%%%%%%%%%%%%%%%%%%%%%%%%%%%%%%%%%%%%%%%%%%%%%%%%%%%%%%%%%%%%%%%%%%%%%%%%%%%
%%%%%%%%%%%%%%%%%%%%%%%%%%%%%%%%%%%%%%%%%%%%%%%%%%%%%%%%%%%%%%%%%%%%%%%%%%%%%%%%
\begin{table}[!h]
\tabcolsep 0pt \caption{Numerical check of Maxwell equal area law
for $P-r_+$ graph for $m=0.5, \alpha=0.8/6, \beta=0.5,
c=c_1=c_2=2, c_3=0.2, c_4=-0.2, d=6, q=1$ } \vspace*{-12pt}
\begin{center}
\def\temptablewidth{1\textwidth}
{\rule{\temptablewidth}{2pt}}
\begin{tabular*}{\temptablewidth}{@{\extracolsep{\fill}}cccccccc}
$T_*$ & $P_*$ & $r_1$ &$r_2$ &$r_3$ &$P_*(r_3-r_1)$ &
$\int^{r_3}_{r_1}Pdr_+$ & relative error \\   \hline
    0.307158782    & 0.019064415 $(0.5P_c)$     & 1.300131202     &    4.377441744 & 6.821914504 & 0.105269568 & 0.069352210 & 0.517898 \\
    0.325263118 & 0.022877298 $(0.6P_c)$    & 1.416781872    &     3.915076926 & 5.952121403  & 0.103756314 & 0.084296053 & 0.230856 \\
 0.340839656    &  0.026690181 $(0.7P_c)$      &  1.553413113       & 3.546020766       &5.217265181 & 0.097788875 & 0.088679752 & $0.102719$ \\
  0.354238750   & 0.030503064  $(0.8P_c)$   & 1.726021434      &    3.239493661 & 4.550601329 & 0.086158341 & 0.082939703 & $0.038807$
       \end{tabular*}
       {\rule{\temptablewidth}{2pt}}
       \end{center}
       \label{tb707}
       \end{table}

   \begin{table}[!h]
\tabcolsep 0pt \caption{Numerical check of Maxwell equal area law
for $P-V$ graph for $m=0.2, \alpha=0.5, \beta=0.5, c=c_1=c_2=2,
c_3=0.2, c_4=-0.2, d=6, \Lambda=-0.1$ } \vspace*{-12pt}
\begin{center}
\def\temptablewidth{1\textwidth}
{\rule{\temptablewidth}{2pt}}
\begin{tabular*}{\temptablewidth}{@{\extracolsep{\fill}}cccccccc}
$T$ & $P_*$ & $V_1$ &$V_2$ &$V_3$ &$P_*(V_3-V_1)$ &
$\int^{V_3}_{V_1}PdV$ & relative error \\   \hline
    $0.307158782$    & 0.019064415 $(0.5P_c)$    & 19.553945    &     8460.584140 & 77773.176052 & 1482.327320 & 1482.327341 & $1.41669\times10^{-8}$ \\
     0.325263118 & 0.022877298 $(0.6P_c)$   & 30.047852   &    4841.723366 & 39323.971901  & 898.938810 & 898.938820 & $1.11242\times10^{-8}$ \\
 0.340839656    &  0.026690181 $(0.7P_c)$    & 47.613851       & 2951.242158      &20347.592453 & 541.810103 & 541.810105 & $3.69133\times10^{-9}$ \\
   0.354238750   & 0.030503064  $(0.8P_c)$  & 80.636078     &    1877.950521 & 10271.702205 & 310.858742 & 310.858740 & $6.43379\times10^{-9}$
       \end{tabular*}
       {\rule{\temptablewidth}{2pt}}
       \end{center}
       \label{tb708}
       \end{table}

\section{Critical exponents \label{CritExp}}

\label{Sec6} To characterize the critical behavior near the critical point
in the extended phase space, one usually introduces the following critical
exponents
\begin{eqnarray}
C_V&\propto&|t|^{-\alpha_1},  \label{801} \\
\eta&\propto&|t|^{\beta_1},  \label{802} \\
\kappa_T&\propto&|t|^{-\gamma},  \label{803} \\
|P-P_c|&\propto&|v-v_c|^{\delta}.  \label{804}
\end{eqnarray}

Note that we use the notations $\alpha_1$ and $\beta_1$ instead of
the classical notation $\alpha$ and $\beta$ here because $\alpha$
and $\beta$ already have other meanings in this paper. As can be
seen from the above definitions, the exponents $\alpha_1, \beta_1,
\gamma$ and $\delta$ describe the behavior of specific heat $C_V$,
the order parameter $\eta$, the isothermal compressibility
coefficient $\kappa_T$ and the critical isotherm respectively.

Before calculating the above critical exponents, it would be convenient to
define
\begin{equation}
t=\frac{T}{T_c}-1,\;\;\epsilon=\frac{v}{v_c}-1,\;\;p=\frac{P}{P_c}.
\label{805}
\end{equation}

The equation of state in the extended phase space has been derived in Ref.
\cite{HendiEPGBBIMass} as
\begin{eqnarray}
P&=&\frac{d_2(2\kappa \alpha^{\prime }+r_+^2)T}{4r_+^3}-\frac{%
m^2cd_2[d_3d_4c^2(d_5cc_4+c_3r_+)+r_+^2(d_3cc_2+c_1r_+)]}{16\pi r_+^4}
\notag \\
&\;&+\frac{\beta^2(\sqrt{1+\eta_+}-1)}{4\pi}-\frac{d_2\kappa(d_5\kappa
\alpha^{\prime }+d_3r_+^2)}{16\pi r_+^4},  \label{806}
\end{eqnarray}
where $\alpha^{\prime }=d_3d_4\alpha, \eta_+=\frac{d_2d_3q^2}{%
2\beta^2r_+^{2d_2}}$. Identifying the specific volume $v$ as $v=\frac{4r_+}{%
d_2}$, the equation of state can be reorganized as
\begin{eqnarray}
P&=&\frac{T}{v}+\frac{32\kappa d_3d_4\alpha T}{d_2^2v^3} -\frac{%
m^2c[16d_3d_4c^2(4d_5cc_4+c_3d_2v)+d_2^2v^2(4d_3cc_2+c_1d_2v)]}{4\pi d_2^3
v^4}  \notag \\
&\;&+\frac{\beta^2\left(\sqrt{1+\frac{d_2d_3q^2 4^{2d_2}} {%
2\beta^2d_2^{2d_2}v^{2d_2}}}-1\right)}{4\pi}-\frac{\kappa(16d_5\kappa
d_3d_4\alpha+d_3d_2^2v^2)}{\pi d_2^3v^4}.  \label{807}
\end{eqnarray}

Then the equation of state in the extended phase space can be expanded as
\begin{equation}
p=1+p_{10}t+p_{01}\epsilon+p_{11}t\epsilon+p_{02}\epsilon^2+
p_{03}\epsilon^3+O(t\epsilon^2,\epsilon^4).  \label{808}
\end{equation}
where the expansion coefficients can be calculated as
\begin{eqnarray}
p_{01}&=&p_{02}=0,  \label{809} \\
p_{10}&=&\frac{T_c}{v_cP_c}+\frac{32d_3d_4\alpha \kappa T_c}{d_2^2v_c^3P_c},
\label{810} \\
p_{11}&=&-\frac{T_c}{v_cP_c}-\frac{96d_3d_4\alpha \kappa T_c}{d_2^2v_c^3P_c},
\label{811} \\
p_{03}&=&-\frac{T_c}{v_cP_c}+\frac{4d_3\kappa
[d_2v_c(d_2v_c-80d_4T_c\alpha)+80d_4d_5\alpha\kappa]}{P_cd_2^3v_c^4\pi}
\notag \\
&\;&+\frac{cm^2(1280c^3c_4d_3d_4d_5+160c^2c_3d_2d_3d_4v_c+
16cc_2d_2^2d_3v_c^2+c_1d_2^3v_c^3)}{4P_cd_2^3v_c^4\pi}  \notag \\
&\;&-[256^{d_2}d_2^{2-2d_2}(1+d_2)(2+d_2)d_3^2q^4+2^{1+4d_2}
d_2(4+d_2)(1+2d_2)d_3q^2v_c^{2d_2}%
\beta^2+8d_2^{2d_2}(1+d_2)(1+2d_2)v_c^{4d_2}\beta^4]  \notag \\
&\;&\times \frac{2^{\frac{1}{2}+4d_2}d_2^2d_3q^2v_c^{-2d_2}}{48P_c\pi \sqrt{%
2+\frac{16^{d_2}d_2^{1-2d_2}d_3q^2v_c^{-2d_2}}{\beta^2}}
(16^{d_2}d_2d_3q^2+2d_2^{2d_2}v_c^{2d_2}\beta^2)^2} .  \label{812}
\end{eqnarray}

From the equal area law, one can further derive
\begin{equation}
\int^{\epsilon_s}_{\epsilon_l}\epsilon \frac{d p} {d\epsilon}d \epsilon=0,
\label{813}
\end{equation}
where $\frac{d p}{d\epsilon}$ can be calculated as $p_{11}t+3p_{03}%
\epsilon^2 $. Denoting the subscript "$l$" and "$s$" as the
quantity of large black hole and small black hole, respectively,
one can obtain
\begin{equation}
p_{11}t(\epsilon^2_s-\epsilon^2_l)+\frac{3}{2} p_{03}(\epsilon^4_s-%
\epsilon^4_l)=0.  \label{814}
\end{equation}

On the other hand, the pressure of large black hole equals to that of small
black hole as follow
\begin{equation}
1+p_{10}t+p_{11}t\epsilon_l+p_{03}\epsilon_l^3=1+p_{10}t+p_{11}t%
\epsilon_s+p_{03}\epsilon_s^3,  \label{815}
\end{equation}
because during the phase transition the pressure of the black hole keeps
unchanged.

With Eqs. (\ref{814}) and (\ref{815}), one can get
\begin{equation}
\epsilon_l=-\epsilon_s=\sqrt{\frac{-p_{11}t}{p_{03}}}.  \label{816}
\end{equation}

So the order parameter can be derived as
\begin{equation}
\eta=v_l-v_s=v_c(\epsilon_l-\epsilon_s)=2v_c\epsilon_l\propto\sqrt{-t},
\label{817}
\end{equation}
leading to the conclusion that $\beta_1=1/2$.

It is not difficult to deduce that
\begin{equation}
\kappa_T=\left.-\frac{1}{v}\frac{\partial v}{\partial P} \right|_{v_c}%
\propto \left.-\frac{1}{\frac{\partial p}{\partial \epsilon}}%
\right|_{\epsilon=0}=-\frac{1}{p_{11}t},  \label{818}
\end{equation}
with which one can draw the conclusion that $\gamma=1$.

One can obtain the critical isotherm by substituting $t=0$ into Eq. (\ref%
{808})
\begin{equation}
p-1=p_{03}\epsilon^3,  \label{819}
\end{equation}
implying that $\delta=3$.

The entropy $S$ does not depend on the Hawking temperature $T$. So
the specific heat with fixed volume $C_V$ is equal to zero, with
the critical exponent $\alpha_1=0$.

The above exponents are totally the same as those in former literature. It
can be attributed to the effect of mean field theory.

\section{Analytical check of Ehrenfest equations at the critical point in
the extended phase space \label{Ehrenfest}}

It is important to classify the nature of phase transition. As we
know, the Clausius-Clapeyron equation is satisfied for a first
order phase transition while for a second order phase transition
one can utilize the famous Ehrenfest equations as follows
\begin{eqnarray}
(\frac{\partial P}{\partial T})_S&=&\frac{C_{P_2}-C_{P_1}}{VT(\tilde{\alpha}%
_2-\tilde{\alpha}_1)}=\frac{\Delta C_P}{VT\Delta \tilde{\alpha}},
\label{901} \\
(\frac{\partial P}{\partial T})_V&=&\frac{\tilde{\alpha}_2-\tilde{\alpha}_1}{%
\kappa_{T_2}-\kappa_{T_1}}=\frac{\Delta \tilde{\alpha}}{\Delta\kappa_T},
\label{902}
\end{eqnarray}%
where the volume expansion coefficient $\tilde{\alpha}=\frac{1}{V} (\frac{%
\partial V}{\partial T})_P$ and isothermal compressibility coefficient $%
\kappa_T=-\frac{1}{V}(\frac{\partial V}{\partial P})_T$. Note that we use
the notation $\tilde{\alpha}$ instead of the classical notation $\alpha$
here because $\alpha$ already has other meaning in this paper.

Utilizing the definition of $\tilde{\alpha}$, one can derive
\begin{equation}
V\tilde{\alpha}=(\frac{\partial V}{\partial T})_P=(\frac{\partial V}{%
\partial S})_P(\frac{\partial S}{\partial T})_P=(\frac{\partial V}{\partial S%
})_P(\frac{C_P }{T}).  \label{903}
\end{equation}

So the R.H.S of Eq. (\ref{901}) can be obtained as
\begin{equation}
\frac{\Delta C_P}{TV\Delta \tilde{\alpha}}=[(\frac{\partial S}{\partial V}%
)_P]_c.  \label{904}
\end{equation}

The subscript "c" here denotes the corresponding quantity at the critical
point. It is not difficult to obtain
\begin{equation}
\frac{\Delta C_P}{TV\Delta \tilde{\alpha}}=\frac{d_2(r_c^2+2d_4d_3\alpha
\kappa)}{4r_c^3}.  \label{905}
\end{equation}

Utilizing Eq. (\ref{806}), the L.H.S of Eq. (\ref{901}) can be derived as
\begin{equation}
[(\frac{\partial P}{\partial T})_S]_c=\frac{d_2(r_c^2+2d_4d_3\alpha \kappa)}{%
4r_c^3}.  \label{906}
\end{equation}

From Eqs. (\ref{905}) and (\ref{906}), we can draw the conclusion that the
first equation of Ehrenfest equations is valid at the critical point.

The L.H.S of Eq. (\ref{902}) can be obtained as
\begin{equation}
[(\frac{\partial P}{\partial T})_V]_c=\frac{d_2(r_c^2+2d_4d_3\alpha \kappa)}{%
4r_c^3}.  \label{907}
\end{equation}

With both the definitions of $\kappa_T$ and $\tilde{\alpha}$, one can deduce
\begin{equation}
V\kappa_T=-(\frac{\partial V}{\partial P})_T=(\frac{\partial T}{\partial P}%
)_V(\frac{\partial V}{\partial T})_P=(\frac{\partial T}{\partial P})_VV%
\tilde{\alpha},  \label{908}
\end{equation}%
from which we can calculate the R.H.S of Eq. (\ref{902}) and get
\begin{equation}
\frac{\Delta \tilde{\alpha}}{\Delta \kappa_T}=[(\frac{\partial P}{\partial T}%
)_V]_c=\frac{d_2(r_c^2+2d_4d_3\alpha \kappa)}{4r_c^3}.  \label{909}
\end{equation}

In the derivation of Eq. (\ref{908}), we have utilized the thermodynamic
identity $(\frac{\partial V}{\partial P})_T(\frac{\partial T}{\partial V})_P(%
\frac{\partial P}{\partial T})_V=-1$. Eq. (\ref{909}) reveals the validity
of the second equation of Ehrenfest equations. With Eqs. (\ref{905}) and (%
\ref{909}), the Prigogine-Defay (PD) ratio can be calculated as
\begin{equation}
\Pi=\frac{\Delta C_P \Delta \kappa_T}{TV(\Delta \tilde{\alpha})^2}=1.
\label{910}
\end{equation}

The above equation and the validity of Ehrenfest equations show that
GB-BI-massive black holes undergo second order phase transition at the
critical point of $P-V$ criticality in the extended phase space. The result
here is consistent with the nature of liquid-gas phase transition at the
critical point and support the findings in former literatures \cite%
{xiong1,xiong2,xiong4}.

\section{Closing Remarks}

In this paper, we have studied thermodynamical behavior of
Einstein-GB-massive black holes in the presence of BI nonlinear
electromagnetic field near critical point.

First, some comments regarding the effects of mass of graviton,
nonlinearity of the electromagnetic field and power of gravity
(value of the GB curvature term) on phase structure and its
complexity were given. In addition, geometrical thermodynamics was
used to investigate phase transition of these black holes based on
canonical ensemble.

Next, by using the denominator of heat capacity and the
proportionality between the cosmological constant and
thermodynamic pressure, critical behavior of these black holes was
investigated. It was shown that these black holes enjoy an anomaly
in their phase structure. In other words, in addition to Van der
Waals like phase transition in their phase diagrams, these black
holes enjoy another type of phase transition which is different
from usual Van der Waals like phase transition. Plotted
coexistence curves also confirmed that only one second order phase
transition exists for these black holes.

Moreover, Maxwell equal area law was employed to investigate the Van der
Waals like behavior and structure of these black holes. It was shown that
Maxwell equal area law holds for $T-r_{+}$, $T-S$ and $P-V$ diagrams while
it fails regarding $P-r_{+}$ curves. Calculations regarding critical
exponent showed that these exponents are independent of massive gravity and
are same as those derived previously. Finally, the Ehrenfest equations were
used to determine the type of phase transition. It was shown that these
black holes undergo second order phase transition at the critical point.

\begin{acknowledgements}
We would like to express our sincere gratitude to both the referee
and the editor whose creative work helps us to improve the quality
of this paper greatly. We thank Shiraz University Research
Council. This work has been supported financially by the Research
Institute for Astronomy and Astrophysics of Maragha, Iran.
Gu-Qiang Li and Jie-Xiong Mo are supported by National Natural
Science Foundation of China (Grant No.11605082), Guangdong Natural
Science Foundation of China (Grant Nos.2016A030307051,
2016A030310363, 2015A030313789) and Department of Education of
Guangdong Province of China (Grant No.2014KQNCX191).
\end{acknowledgements}

\end{document}